\documentclass[10pt,journal,twocolumn,twoside]{autart}
\usepackage{graphicx}          % Include this line if your 
\usepackage{amssymb}  % assumes amsmath package installed
\usepackage{amsmath} % assumes amsmath package installed
\usepackage{algorithm}%,algorithmic}
\usepackage{algpseudocode}
\usepackage{subcaption}

\usepackage{hyperref}

\usepackage{color}  
\begin{document}
\begin{frontmatter} 
\title{A nonlinear tracking model predictive control scheme for  dynamic target signals\thanksref{footnoteinfo}}  %unreachable
\thanks[footnoteinfo]{The material in this paper was partially presented at the 6th IFAC Conference on Nonlinear Model Predictive Control, August 19–-22, 2018,  Madison, Wisconsin (USA).}
\author{Johannes K\"ohler$^1$}\ead{johannes.koehler@ist.uni-stuttgart.de}, %
\author{Matthias A. M\"uller$^2$}\ead{mueller@irt.uni-hannover.de}, %
\author{Frank Allg\"ower$^1$}\ead{frank.allgower@ist.uni-stuttgart.de}%
\address{$^1$Institute for Systems Theory and Automatic Control, University of Stuttgart, 70550 Stuttgart, Germany} 
\address{$^2$Institute of Automatic Control,Leibniz University Hannover, 30167 Hannover, Germany} 
\date{ \copyright  2020 Elsevier Ltd.  All rights reserved.}
                             % keyword list or with the 
\begin{abstract}                          % Abstract of not more than 200 words. (120words) 
We present a nonlinear model predictive control (MPC) scheme for tracking of dynamic target signals. 
The scheme combines stabilization and dynamic trajectory planning in one layer, thus ensuring constraint satisfaction irrespective of changes in the dynamic target signal. 
For periodic target signals we ensure exponential stability of the optimal reachable periodic trajectory using suitable terminal ingredients and a convexity condition for the  underlying periodic optimal control problem. 
Furthermore, we introduce an online optimization of the terminal set size to automate the trade-off between fast convergence and operation close to the constraints. 
In addition, we show how stabilization and dynamic trajectory planning can be formulated as partially decoupled optimization problems, which reduces the computational demand while ensuring recursive feasibility and convergence.
The main tool to enable the proposed design is a novel reference generic offline computation that provides suitable terminal ingredients for tracking of dynamic reference trajectories. 
The practicality of this approach is demonstrated on benchmark examples, which demonstrates superior performance compared to state of the art approaches.  
\end{abstract}
\end{frontmatter}

%
%!TEX root = ./PeriodicTracking_Automatica.tex
%%%%%%%%%%%%%%%%%%%%%%%%%%%%%%%%%%%%%%%%%%%%%%%%%%%%%%%%%%%%%%%%%%%%%%%%%%%%%%%
\section{Introduction}
Model Predictive Control (MPC) is a well established control method, that computes the control input by repeatedly solving an optimization problem online~\cite{rawlings2017model}.
The main advantages of MPC are the ability to cope with nonlinear dynamics, hard state and input constraints, and the inclusion of performance criteria. 

\textit{Motivation: }
Most of the existing theoretical results for MPC consider the problem of stabilizing some given steady-state~\cite{rawlings2017model}. 
Theoretical properties, such as recursive feasibility and asymptotic stability can be ensured by including suitable terminal ingredients (terminal set and terminal cost) in the optimization problem, which are computed offline, compare~\cite{chen1998quasi,mayne2000constrained}. 

In many applications, the control goal goes beyond the stabilization of a pre-determined setpoint. 
These practical challenges include tracking of changing reference setpoints, stabilization of dynamic trajectories, output regulation and general economic optimal operation. 
Designing MPC schemes that provide theoretical guarantees (recursive feasibility, stability and performance) for such control problems is the focus of much research.

\textit{Related work: }
In~\cite{faulwasser2012optimization,JK_QINF} tracking of \textit{reachable} dynamic reference trajectories  is studied and stability is ensured by using suitable terminal ingredients. 
In~\cite{kohlernonlinear19}, a tracking MPC scheme without terminal ingredients for \textit{unreachable} target signals is studied and (practical) stability of the optimal \textit{reachable} trajectory is established. 
The MPC scheme is simple to implement, however, the theoretical guarantees depend on a sufficiently large prediction horizon that may be conservative for many applications. 

A promising alternative to tackle the problem of \textit{unreachable} target signals, is the simultaneous optimization of an artificial reference, which is pursued in~\cite{limon2008mpc,limon2018nonlinear,limon2016mpc,fagiano2013generalized,muller2013economic,ferramosca2014economic}.  %
By using terminal constraints for the artificial reference, these schemes provide a large region of attraction and ensure recursive feasibility independent of the (typically exogenous) target signal. 
In particular, in~\cite{limon2008mpc} a setpoint tracking MPC scheme for linear systems has been introduced based on simultaneous optimization of an artificial steady state and tracking of this steady state. 
Compared to a standard MPC formulation, this scheme ensures recursive feasibility independent of the target signal and provides a large region of attraction. 
In~\cite{simon2014reference}, the complexity of the polyhedral invariant set for tracking from~\cite{limon2008mpc} has been reduced by considering an online optimization of the terminal set size. 
In~\cite{limon2016mpc}, for linear systems the approach in~\cite{limon2008mpc} has been extended to periodic target signals by using an artificial \textit{periodic} trajectory and a terminal equality constraint. 
Similarly, in~\cite{limon2014single} for linear systems the \textit{economically} optimal periodic trajectory is stabilized using a tracking formulation. 
 The method in~\cite{limon2008mpc} has been extended to setpoint tracking for \textit{nonlinear} systems in~\cite{limon2018nonlinear}.
\textit{Economic} MPC schemes based on artificial steady states have been considered in~\cite{fagiano2013generalized,muller2013economic,ferramosca2014economic}, for both linear and nonlinear systems.

\textit{Contribution: } 
The goal of this paper is to generalize and unify the methodologies from~\cite{limon2008mpc,limon2018nonlinear,limon2016mpc} to design nonlinear MPC schemes that exponentially stabilize the optimal \textit{reachable} periodic trajectory given a possibly \textit{unreachable} periodic output target signal, which is a fundamental step towards practical nonlinear MPC schemes. 
We first generalize the conditions on the terminal ingredients, such that we can present a unified theorem that incompases terminal equality constraints (TEC) (under suitable controllabiltiy conditions) and suitable terminal cost and terminal set (which can be designed using the \textit{reference generic} offline computation in~\cite{JK_QINF}). 
Then we design a nonlinear tracking MPC scheme for \textit{unreachable} periodic target signals. 
We provide a novel proof to show that the optimal \textit{reachable} periodic trajectory is exponentially stable for the resulting closed-loop system, if the set of feasible periodic output trajectories is convex. 
Furthermore, we extend this method to allow for an online optimization of the terminal set size, which significantly improves the performance, similar to~\cite{simon2014reference}. 
This extension requires one additional scalar optimization variable and automates the trade-off between fast convergence and operation close to the constraints, which typically needs to be decided offline.
In addition, we provide a novel algorithm that partially decouples the reference trajectory updates and the computation of the closed-loop input.
Finally, we demonstrate the applicability and practicality of the proposed methodology using two nonlinear benchmark examples. 
Furthermore, we showcase superior performance compared to state of the art approaches in a quantitative comparison by including the proposed terminal ingredients and optimizing terminal set size online.  

A preliminary version of the proposed approach can be found in the conference
proceedings~\cite{kohler2018mpc}.
Compared to the conference version, we provide a more comprehensive exposition of the subject, unify the consideration of different terminal ingredients, include an online optimization of the terminal set size, discuss how to partially decouple the reference update and add an example with a quantitative comparison to state of the art approaches.

\textit{Outline: }
Sec.~\ref{sec:MPC} discusses preliminaries regarding tracking MPC. 
Sec.~\ref{sec:ext} contains the proposed MPC scheme, including theoretical analysis and extensions. 
Sec.~\ref{sec:num} shows the applicability and advantages of the proposed method using numerical examples.
Sec.~\ref{sec:sum} concludes the paper. 

\textit{Notation: }
The quadratic norm with respect to a positive definite matrix $Q=Q^\top$ is denoted by $\|x\|_Q^2=x^\top Q x$. 
By $\mathcal{K}_{\infty}$ we denote the class of functions $\alpha:\mathbb{R}_{\geq 0}\rightarrow\mathbb{R}_{\geq 0}$, which are continuous, strictly increasing, unbounded and satisfy $\alpha(0)=0$. 
The identity matrix is $I_n\in\mathbb{R}^{n\times n}$. 
The interior of a set $\mathcal{X}$ is denoted by $\text{int}(\mathcal{X})$.

%!TEX root = ./PeriodicTracking_Automatica.tex
%%%%%%%%%%%%%%%%%%%%%%%%%%%%%%%%%%%%%%%%%%%%%%%%%%%%%%%%%%%%%%%%%%%%%%%%%%%%%%%
\section{Tracking MPC for reachable references}
\label{sec:MPC}
%!TEX root = ./PeriodicTracking_Automatica.tex
%%%%%%%%%%%%%%%%%%%%%%%%%%%%%%%%%%%%%%%%%%%%%%%%%%%%%%%%%%%%%%%%%%%%%%%%%%%%%%%
We consider the following nonlinear discrete-time system 
$x_{t+1}=f(x_t,u_t)$
 with state $x\in\mathbb{R}^n$, control input $u\in\mathbb{R}^m$, and time $t\in\mathbb{N}$.
We impose point-wise in time constraints on the state and input
$(x_t,u_t)\in \mathcal{Z}$,
with some compact set $\mathcal{Z}$. 
As a preliminary problem, we consider tracking of a given \textit{reachable} reference trajectory $r_t=(x^r_{t},u^r_{t})\in\mathbb{R}^{n+m}$, analogous to~\cite{JK_QINF}. 
\begin{assum}
\label{ass:ref}
The reference trajectory $r$ satisfies $r_t\in\mathcal{Z}_r$, for all $t\geq 0$, with some set $\mathcal{Z}_r\subseteq\text{int}(\mathcal{Z})$. 
Furthermore, the evolution of the reference trajectory is restricted by $r_{t+1}\in\mathcal{R}(r_t)$, with $\mathcal{R}(r)=\{(x^{r+},u^{r+})\in\mathcal{Z}_r|~x^{r+}=f(x^r,u^r)\}$. 
\end{assum}
This assumption characterizes that the reference trajectory $r$ is reachable, i.e., follows the dynamics $f$ and (strictly) satisfies the constraints $\mathcal{Z}$ for all times. 
Denote the tracking error by $e_{t}:=x_t-x^r_{t}$. 
The control goal is stability of the tracking error $e_{t}=0$ and constraint satisfaction $(x_t,u_t)\in\mathcal{Z}$, $\forall t\geq 0$.   
We define the quadratic reference tracking stage cost 
\begin{align*}
\ell(x,u,r)=\|x-x^{r}\|_Q^2+\|u-u^{r}\|_R^2,
\end{align*}
with positive definite weighting matrices $Q,~R$.
Denote the reference $r$ over the prediction horizon $N$ by  ${r}_{\cdot|t}$ with $r_{k|t}=r_{t+k}$, $k=0,\dots,N$. 
Given a predicted state and input sequence $x_{\cdot|t}\in\mathbb{R}^{n\times (N+1) }$, $u_{\cdot|t}\in\mathbb{R}^{m\times N}$ the tracking cost with respect to the reference $r_{\cdot|t}$ is given by
\begin{align*}
J_N(x_{\cdot|t},u_{\cdot|t},r_{\cdot|t}):=&\sum_{k=0}^{N-1}\ell(x_{k|t},u_{k|t},r_{k|t})
+V_f(x_{N|t},r_{N|t}),
\end{align*}
with the terminal cost $V_f$. 
The MPC scheme is based on the following (standard) MPC optimization problem
\begin{subequations}
\label{eq:MPC}
\begin{align}
\label{eq:MPC_cost}
V_t:=&\min_{u_{\cdot|t}}J_N(x_{\cdot|t},u_{\cdot|t},r_{\cdot|t})\\
\label{eq:MPC_dyn}
\text{s.t. }&x_{k+1|t}=f(x_{k|t},u_{k|t}),\\
&(x_{k|t},u_{k|t})\in\mathcal{Z},~k=0,\dots,N-1,\\
&x_{0|t}=x_t,~x_{N|t}\in\mathcal{X}_f({r}_{N|t}),
\end{align}
\end{subequations}
where $\mathcal{X}_f(r)\subset\mathbb{R}^n$ denotes the terminal set. 
The solution to this optimization problem are the value function $V_t$ and the optimal state and input trajectories $x^*_{\cdot|t}$, $u^*_{\cdot|t}$. 
For simplicity, we assume throughout the paper that the optimization problems admit a unique minimizer\footnote{%
Existence of a minimizer is, e.g., guaranteed if $f,\ell,V_f$ are continuous and the constraint set is compact. If the minimizer is not unique, an arbitrary minimizer can be chosen.}. 
In closed-loop operation we apply the first part of the optimized input trajectory to the system, leading to the closed-loop system 
$x_{t+1}=f(x_t,u^*_{0|t})=x^*_{1|t}$, $t\geq 0$. 
As discussed in the introduction, we need suitable terminal ingredients to ensure stability and recursive feasibility for the closed loop. 
\begin{assum}
\label{ass:term_gen}
There exist a terminal controller $k_f:\mathbb{R}^{n}\times\mathcal{Z}_r\rightarrow\mathbb{R}^m$, a terminal cost $V_f:\mathbb{R}^{n}\times\mathcal{Z}_r$ and a terminal set $\mathcal{X}_f(r)\subset\mathbb{R}^n$, such that the following properties hold for any $r\in\mathcal{Z}_r$, any $x\in\mathcal{X}_f(r)$ and any $r^+\in\mathcal{R}(r)$
\begin{subequations}
\label{eq:term_gen}
\begin{align}
\label{eq:term_dec}
V_f(x^+,r^+)\leq& V_f(x,r)-\ell(x,k_f(x,r),r),\\
\label{eq:term_con}
(x,k_f(x,r))\in&\mathcal{Z},\\
\label{eq:term_invariant}
x^+\in&\mathcal{X}_f(r^+),
\end{align}
\end{subequations}
with $x^+=f(x,k_f(x,r))$. 
Furthermore, there exist constants $c_u$, $\epsilon>0$, such that for any reference $r_{\cdot|t}$ satisfying Ass.~\ref{ass:ref}, and any  $x_t\in\mathbb{R}^n$ with $\|e_{t}\|_Q\leq \epsilon$, Problem~\eqref{eq:MPC} is feasible and the value function  satisfies
\begin{align}
\label{eq:value_quad_bound}
V_t\leq c_u\|e_{t}\|_Q^2.
\end{align}
\end{assum}
The first set of conditions~\eqref{eq:term_gen} is standard in (reference tracking) MPC, compare for example~\cite{chen1998quasi,rawlings2017model,faulwasser2012optimization,JK_QINF}. 
Condition~\eqref{eq:value_quad_bound} ensures that the value function admits a local quadratic upper bound.\footnote{The \textit{weak controllability condition}~\cite[Ass.~2.17]{rawlings2017model}, which does not require feasibility of Problem~\eqref{eq:MPC} locally around the reference trajectory $r$, would be sufficient to prove Theorem~\ref{thm:MPC}. 
Assumption~\ref{ass:term_gen}, similar to condition~\cite[Ass.~4]{limon2018nonlinear} used for setpoint stabilization, is stronger and will be used in the proof of Theorem~\ref{thm:track} in Section~\ref{sec:ext}. }
%~\cite[Ass.~2.23]{rawlings2009model}
In Prop.~\ref{prop:TEC} and Lemma~\ref{lemma:LPV_term} below, we provide sufficient conditions for Assumption~\ref{ass:term_gen} using controllability and stabilizability conditions. 
\begin{thm}
\label{thm:MPC}
Let Ass.~\ref{ass:ref}--\ref{ass:term_gen} hold. 
Assume that Problem~\eqref{eq:MPC} is feasible at $t=0$. 
Then Problem~\eqref{eq:MPC} is recursively feasible and the tracking error $e=0$ is uniformly exponentially stable for the resulting closed-loop system.
\end{thm}
\begin{pf}
This theorem is a straight forward extension of standard MPC results~\cite{rawlings2017model}, compare also~\cite{JK_QINF,faulwasser2012optimization}. 
Given the optimal solution $u^*_{\cdot|t}$, the candidate sequence 
\begin{align*}
u_{k|t+1}=   \begin{cases}
u^*_{k+1|t}& 0\leq k\leq N-2  \\
k_f(x^*_{N|t},r_{N|t})&k=N-1
\end{cases},
\end{align*}
 is a feasible solution to~\eqref{eq:MPC} and implies 
\begin{align}
\label{eq:V}
&V_{t+1}- V_t\leq-\ell(x_t,u_t,r_t)\leq -\|e_{t}\|_Q^2. 
\end{align}
Compactness of $\mathcal{Z}$ in combination with the local quadratic upper bound~\eqref{eq:value_quad_bound} imply 
$\|e_{t}\|_Q^2\leq V_t\leq c_v \|e_{t}\|_Q^2$,
for some $c_v\geq c_u\geq 1$ and all $x_t$ such that Problem~\eqref{eq:MPC} is feasible, compare~\cite[Prop.~2.16]{rawlings2017model}.
%\cite[Prop.~2.18]{rawlings2012postface}.
Uniform exponential stability follows from standard Lyapunov arguments based on the value function $V$. $\hfill\square$
\end{pf}
Thus, the closed-loop tracking MPC scheme given by \eqref{eq:MPC} has all the (standard) desirable properties, in case that a reachable reference trajectory shall be tracked. 
The main contribution of this work (see Sec.~\ref{sec:ext}) is the development of MPC schemes for more general tracking problems, including cases where an unreachable dynamic target signal is given. 
We note that the tracking MPC scheme~\eqref{eq:MPC} (and correspondingly also the proposed scheme in Section~\ref{sec:ext}) can be easily modified to ensure \textit{robust} reference tracking, compare~\cite{Robust_TAC_19}, \cite[Thm.~2]{JK_QINF}, and the discussion in Remark~\ref{rk:robust}.

In the following, we briefly detail how Assumption~\ref{ass:term_gen} can be satisfied using either a terminal equality constraint or a suitable terminal cost. 
%TEC
%!TEX root = ./PeriodicTracking_Automatica.tex
%%%%%%%%%%%%%%%%%%%%%%%%%%%%%%%%%%%%%%%%%%%%%%%%%%%%%%%%%%%%%%%%%%%%%%%%%%%%%%%
\subsubsection*{Terminal equality constraint - controllability}
The following proposition shows that a simple  terminal equality constraint (TEC) with $\mathcal{X}_f(r)=x_r$, $V_f=0$  satisfies Ass.~\ref{ass:term_gen}, if the system is locally uniformly exponentially finite time controllable.  
\begin{prop}
\label{prop:TEC}
Suppose there exist constants $N_0\in\mathbb{N}$, $c_u$, $\epsilon_{\text{TEC}}>0$, such that for any reference trajectory $r_{\cdot|t}$ satisfying Ass.~\ref{ass:ref}, and any state $x_t\in\mathbb{R}^n$ satisfying $\|e_{t}\|_Q\leq {\epsilon}_{\text{TEC}}$, there exists an input trajectory $u_{\cdot|t}\in\mathbb{R}^{m\times N_0}$ such that
\begin{align*}
&x_{k+1|t}=f(x_{k|t},u_{k|t}),\quad x_{N_0|t}=x_{r,N_0|t},\\ 
&J_{N_0}(x_{\cdot|t},u_{\cdot|t},r_{\cdot|t})\leq c_u \|e_{t}\|_Q^2.  
\end{align*}
Then for any $N\geq N_0$, Ass.~\ref{ass:term_gen} holds with $\mathcal{X}_f(r)=x^r$, $V_f(x,r)=0$, $k_f(x,r)=u^r$ and some constant $\epsilon>0$. 
\end{prop}
\begin{pf}
Satisfaction of~\eqref{eq:term_gen} is standard. 
Consider Problem~\eqref{eq:MPC} at time $t$ with initial condition satisfying $\|e_{t}\|_Q\leq \epsilon\leq \epsilon_{\text{TEC}}$, some prediction horizon $N\geq N_0$, and the input $u_{\cdot|t}$ appended by $u_{k|t}=u^r_{k|t}$,~$k\geq N_0$.
This candidate input $u$  satisfies 
$J_N(x_{\cdot|t},u_{\cdot|t},r_{\cdot|t})\leq c_u\|e_{t}\|_Q^2$
and  $\|x_{k|t}-x^r_{k|t}\|_Q^2+\|u_{k|t}-u^r_{k|t}\|_R^2\leq c_u\epsilon^2$. 
Given that $r_{k|t}\in\mathcal{Z}_r\subseteq\text{int}(\mathcal{Z})$, there exists a small enough constant $\epsilon\in(0,{\epsilon}_{\text{TEC}}]$, such that $(x_{k|t},u_{k|t})\in\mathcal{Z}$. 
Thus, the candidate solution is a feasible solution to~\eqref{eq:MPC} and the bound~\eqref{eq:value_quad_bound} holds. 
$\hfill\square$
\end{pf}
The considered controllability condition holds, for example, if the dynamic $f$ is continuously differentiable and the linearized dynamic $f$ around any reachable trajectory $r$ is $N_0$-step uniformly controllable. 
%, compare also~\cite[Ass.~10]{muller2016economic}.\cite[Assumption 2.17]{rawlings2017model}

%Terminal cost
%!TEX root = ./PeriodicTracking_Automatica.tex
%%%%%%%%%%%%%%%%%%%%%%%%%%%%%%%%%%%%%%%%%%%%%%%%%%%%%%%%%%%%%%%%%%%%%%%%%%%%%%%
%
\subsubsection*{Terminal cost - reference generic offline computations}
In the following, we discuss how to compute a terminal cost $V_f$ and a terminal set $\mathcal{X}_f$ satisfying Ass.~\ref{ass:term_gen}.
Denote the Jacobian of $f$ evaluated around an arbitrary point $r\in\mathcal{Z}_r$ by
\begin{align}
\label{eq:A_r}
A(r)=\left.\left[\dfrac{\partial f}{\partial x}\right]\right|_{(x,u)=r},\quad B(r)=\left.\left[\dfrac{\partial f}{\partial u}\right]\right|_{(x,u)=r}. 
\end{align}
\begin{lem}
\label{lemma:LPV_term}
Suppose that $f$ is twice continuously differentiable. 
Assume that there exist a matrix $K_f(r)\in\mathbb{R}^{m\times n}$ and a positive definite matrix $P_f(r)\in\mathbb{R}^{n\times n}$ continuous in $r$, such that for any  $r\in\mathcal{Z}_r$, $r^+\in\mathcal{R}(r)$, the following matrix inequality is satisfied
\begin{align}
\label{eq:lpv}
&(A(r)+B(r)K_f(r))^\top P_f(r^+)(A(r)+B(r)K_f(r))\nonumber\\
\leq& P_f(r)-(Q+K_f(r)^\top R K_f(r))-\tilde{\epsilon} I_n
\end{align}
 with some constant $\tilde{\epsilon}>0$. 
Then there exists a sufficiently small $\alpha$, such that Ass.~\ref{ass:term_gen} is satisfied for any $N\geq 0$ with 
\begin{align*}
&V_f(x,r)=\|x-x^r\|_{P_f(r)}^2,~ c_u=\sup_{r\in\mathcal{Z}_r}\lambda_{\max}(P_f(r),Q), \\
&\mathcal{X}_f(r)=\{x\in\mathbb{R}^n|~V_f(x,r)\leq \alpha\},~\epsilon=\sqrt{\alpha/c_u}, \\
&k_f(x,r)=u^r+K_f(r)\cdot (x-x^r),
\end{align*}
where $\lambda_{\max}(P,Q)$ denotes the maximal generalized eigenvalue solving $(P-\lambda Q)v=0$, for some $v\in\mathbb{R}^n$.
\end{lem}
\begin{pf}
Satisfaction of conditions~\eqref{eq:term_gen} follows from~\cite[Lemma~1]{JK_QINF} based on standard differentiability/Lipschitz continuity arguments, compare also~\cite{chen1998quasi,rawlings2017model,faulwasser2012optimization}.  
Furthermore, $\|x-x^r\|_Q^2\leq \epsilon$ implies $V_f(x,r)\leq c_u\|x-x^r\|_Q^2\leq \alpha$ and hence $x\in\mathcal{X}_f(r)$. 
Using standard arguments (compare for example~\cite[Prop.~2.35]{rawlings2017model}), within the terminal set $\mathcal{X}_f$ the terminal controller $k_f$ is a feasible solution to~\eqref{eq:MPC} and thus using~\eqref{eq:term_dec} the terminal cost $V_f$ is an upper bound on the value function $V$, which ensures~\eqref{eq:value_quad_bound}. 
$\hfill\square$
\end{pf}
Proposition~3 in~\cite{JK_QINF} 
provides a semidefinite program (SDP) to compute matrices $P_f,~K_f$ satisfying~\eqref{eq:lpv}  
using a parametrization of the form $P_f=X^{-1}$, $K_f=YP_f$, $\theta_j:\mathcal{Z}\rightarrow \mathbb{R}$, 
\begin{align*}
X(r)=X_0+\sum_{j} \theta_j(r) X_j,~Y(r)=Y_0+\sum_{j} \theta_j(r) Y_j,
\end{align*}
using methods for (quasi) linear parameter varying (LPV) systems and gain scheduling~\cite{rugh2000research}. 
This \textit{reference generic} offline design is only done once for a given system $f$ and class of reference trajectories (Ass.~\ref{ass:ref}) and the resulting terminal ingredients can be applied if the reference $r$ is optimized online (cf. Sec.~\ref{sec:ext}).  
Details on the computation with numerical examples can be found in~\cite{JK_QINF}.

%discuss
\textit{Discussion: }
The main advantage of using a terminal equality constraint (TEC, Prop.~\ref{prop:TEC}) is the fact that no offline design is needed.  
One of the main benefits of using a terminal cost/set (QINF, Lemma~\ref{lemma:LPV_term}) is that the desired properties also hold for an arbitrarily small prediction horizon $N$. 
Furthermore, the values of $c_u$, $\epsilon$ can be computed explicitly and are typically significantly less conservative than the ones obtained when using a terminal equality constraint (TEC, Prop~\ref{prop:TEC}), which impacts the closed-loop convergence rate. 
This impact on closed-loop performance is quantitatively investigated with numerical examples in Section~\ref{sec:num}. 
%!TEX root = ./PeriodicTracking_Automatica.tex
%%%%%%%%%%%%%%%%%%%%%%%%%%%%%%%%%%%%%%%%%%%%%%%%%%%%%%%%%%%%%%%%%%%%%%%%%%%%%%%
%Reference
\section{Nonlinear tracking MPC for dynamic target signals}
\label{sec:ext}
In the following, we design a nonlinear MPC scheme for tracking of exogenous (unreachable) periodic target signals using the terminal ingredients in Assumption~\ref{ass:term_gen}. 
For the resulting closed-loop system, Theorem~\ref{thm:track} establishes exponential stability of the optimal reachable periodic trajectory.
In Section~\ref{sec:limon_alpha} we discuss how online optimization of the terminal set size $\alpha$ can speed up convergence, while at the same time allowing optimal operation (arbitrarily) close to the constraints. 
Algorithm~\ref{alg:asyn} in Section~\ref{sec:partial} provides a simple means to reduce the computational demand by partially decoupling the trajectory planning and the reference tracking problem.
%
%!TEX root = ./PeriodicTracking_Automatica.tex
%%%%%%%%%%%%%%%%%%%%%%%%%%%%%%%%%%%%%%%%%%%%%%%%%%%%%%%%%%%%%%%%%%%%%%%%%%%%%%%
\subsection{Nonlinear periodic tracking  MPC }
\label{sec:limon}
In~\cite{limon2008mpc,limon2018nonlinear,limon2016mpc}, tracking MPC schemes based on simultaneous optimization of an artificial reference have been introduced. 
Compared to a standard reference tracking MPC formulation such as~\eqref{eq:MPC}, these schemes ensure recursive feasibility independent of the (potentially unreachable) target signal and provide a large region of attraction. 
In the following, we extend these methods to \textit{nonlinear periodic} reference tracking using  \textit{general terminal ingredients} (Ass.~\ref{ass:term_gen}, using a terminal equality constraint (TEC, Prop.~\ref{prop:TEC}) or a terminal cost/set (QINF, Lemma~\ref{lemma:LPV_term})).   

We consider a nonlinear  output function $y^r=h(x^r,u^r)\in\mathbb{R}^p$ and assume that at time $t$ an exogenous $T$-periodic target signal $y^e_{\cdot|t}\in\mathbb{R}^{p\times T}$ is given. 
For some $T$-periodic reference trajectory $r_{\cdot|t}=(x^r_{\cdot|t},u^r_{\cdot|t})\in\mathbb{R}^{(n+m)\times T}$, 
the tracking cost w.r.t. this target signal $y^e$ is defined as 
\begin{align*}
J_T(r_{\cdot|t},y^e_{\cdot|t}):=&\sum_{j=0}^{T-1} \|\underbrace{h(x^r_{j|t},u^r_{j|t})}_{={y^r_{j|t}}}-y^e_{j|t}\|_S^2=\|y^r_{\cdot|t}-y^e_{\cdot|t}\|_S^2,  
\end{align*}
with some positive definite weighting matrix $S\in\mathbb{R}^{p\times p}$. 
The objective is to stabilize the reachable (Ass.~\ref{ass:ref}) $T$-periodic reference trajectory  $r^{T*}_{\cdot|t}=(x^{T*}_{\cdot|t},u^{T*}_{\cdot|t})$ that minimizes the distance to the signal $y^e$, which is defined as the minimizer to the following optimization problem 
\begin{align}
\label{eq:opt_ref}
&V_T(y^e_{\cdot|t})=\min_{r_{\cdot|t}} J_T(r_{\cdot|t},y^e_{\cdot|t})\\
\text{s.t. }&r_{j+1|t}\in\mathcal{R}(r_{j|t})\subseteq\mathcal{Z}_r,~  r_{0|t}=r_{T|t},~j=0,\dots,T-1.\nonumber
\end{align}
The corresponding output reference is denoted by $y^{T*}_{\cdot|t}$, with $y^{T*}_{k|t}=h(r^{T*}_{k|t})$.   
In order to find and stabilize this periodic trajectory, we consider the following optimization problem, similar to~\cite{limon2016mpc}
\begin{subequations}
\label{eq:limon}
\begin{align}
W_T(x_t,y^e_{\cdot|t})&=\min_{u_{\cdot|t},r_{\cdot|t}}J_N(x_{\cdot|t},u_{\cdot|t},r_{\cdot|t})
+J_T(r_{\cdot|t},y^e_{\cdot|t})\nonumber\\
\text{s.t. }&x_{k+1|t}=f(x_{k|t},u_{k|t}),~ x_{0|t}=x_t,\\
&(x_{k|t},u_{k|t})\in\mathcal{Z},~ 
x_{N|t}\in\mathcal{X}_f(r_{N|t}),\\
\label{eq:limon_Zr}
&r_{j+1|t}\in\mathcal{R}(r_{j|t})\subseteq\mathcal{Z}_r,~r_{l+T|t}=r_{l|t},\\
&j=0,\dots,T-1,~ k=0,\dots,N-1,\nonumber\\
& l=0,\dots,\max\{0,N-T\}. \nonumber
\end{align}
\end{subequations}
The optimal state and input trajectory is given by $u^*_{\cdot|t}$, $x^*_{\cdot|t}$, with the artificial reference $r^*_{\cdot|t}=(x^{r*}_{\cdot|t},u^{r*}_{\cdot|t})$ and the output $y^{r*}_{k|t}=h(r^*_{k|t})$.  
In closed-loop operation we apply the first part of the optimized input trajectory to the system, leading to the following closed-loop system
\begin{align}
\label{eq:close_limon}
x_{t+1}=f(x_t,u^*_{0|t})=x^*_{1|t},\quad t\geq 0.
\end{align}
The rational behind this optimization problem is to penalize the (standard) tracking cost $J_N$ w.r.t. some artificial periodic reference $r$ together with the distance of the output of this artificial reference to the target signal using $J_T$. 
As we will see later in the theoretical analysis (Thm.~\ref{thm:track}) and the numerical examples (Sec.~\ref{sec:num}), this formulation ensures that the closed loop smoothly tracks the optimal reachable periodic trajectory $x^{T*}$. 

%!TEX root = ./PeriodicTracking_Automatica.tex
%%%%%%%%%%%%%%%%%%%%%%%%%%%%%%%%%%%%%%%%%%%%%%%%%%%%%%%%%%%%%%%%%%%%%%%%%%%%%%%
\subsection{Theoretical analysis}
In the following, we derive the theoretical properties of the closed-loop system based on~\eqref{eq:limon}. 
The following condition is used to ensure exponential stability of the optimal trajectory $x^{T*}$, similar to~\cite[Ass.~2]{limon2016mpc},\cite[Ass.~1-2]{limon2018nonlinear}. 
\begin{assum}
\label{ass:unique_convex}
There exist (unique) locally Lipschitz continuous functions $g_x:\mathbb{R}^{p\times T}\rightarrow\mathbb{R}^{n\times T}$, $g_u:\mathbb{R}^{p\times T}\rightarrow\mathbb{R}^{m\times T}$ such that $g_x(y^r_{\cdot|t})=x^r_{\cdot|t}$, $g_u(y^r_{\cdot|t})=u^r_{\cdot|t}$, for any feasible solution to~\eqref{eq:opt_ref}.  
The set of feasible solutions to~\eqref{eq:opt_ref} is convex\footnote{Given two feasible solutions $r_1,~r_2$ with corresponding outputs $y^{r}_{1},~y^{r}_{2}$, the reference $r_{\cdot|t}=(g_x(y^r_{\cdot|t}),g_u(y^r_{\cdot|t}))$ is a feasible solution to~\eqref{eq:opt_ref} with $y^r_{\cdot|t}=\beta y^{r}_{1}+(1-\beta)y^{r}_{2}$, $\beta\in[0,1]$.} in $y^r_{\cdot|t}$.   
\end{assum}
This assumption implies that~\eqref{eq:opt_ref} is a strictly convex problem and the minimizer $r^{T*}$ is unique.
Thus, for any $y^r\neq y^{T*}$ it is possible to incrementally change $y^r$, such that it remains feasible and the cost $J_T$ decreases.  
Furthermore, due to convexity the directional derivative of $J_T$ at $y^{T*}$ in any feasible direction is non-negative, i.e., for  any reference $r_{\cdot|t}$ that satisfies the constraints in~\eqref{eq:opt_ref}, the corresponding output $y^r_{\cdot|t}$ satisfies  
\begin{align}
\label{eq:convex_grad}
(y^r_{\cdot|t}-y^{T*}_{\cdot|t})^\top \nabla_{y^r}J_T|_{y^r=y^{T*}} \geq 0,
\end{align}
which can be equivalently written as 
\begin{align}
\label{eq:unique}
J_T(r_{\cdot|t},y^e_{\cdot|t})\geq V_T(y^e_{\cdot|t})+\|y^{T*}_{\cdot|t}-y^r_{\cdot|t}\|_S^2. 
\end{align}
\begin{rem}
\label{rk:convex}
Similar to~\cite[Remark~1]{limon2018nonlinear}, existence of $g_x$, $g_u$  can be ensured based on the implicit function theorem, if a rank condition on the linearization of a suitably defined $T$-step system  is satisfied and $f$, $h$ are continuously differentiable. \\
For a linear output $h(x,u)=Cx+Du$, the convexity assumption is ensured if the constraint set in~\eqref{eq:opt_ref} (describing reachable periodic orbits) is convex. 
For $T=1$, this reduces to convexity of the steady-state manifold $\mathcal{Z}_r$, which is often easy to verify. 
Furthermore, even in case the set of reachable periodic trajectories is non-convex, it may be possible to choose a suitable nonlinear output $h$, such that the convexity condition (Ass.~\ref{ass:unique_convex}) is satisfied, compare~\cite{cotorruelo2018tracking}. \\
If the convexity condition in Assumption~\ref{ass:unique_convex} is not satisfied, the MPC scheme will not necessarily stabilize the optimal reachable trajectory $x^{T*}$, but could instead stabilize a suboptimal periodic  trajectory, similar to~\cite{muller2013economic}. 
The main alternative to using a tracking scheme with simultaneous optimization of the artificial trajectory such as~\eqref{eq:limon}, would be to directly solve~\eqref{eq:opt_ref} and then apply a tracking MPC for this reachable reference trajectory, compare Section~\ref{sec:MPC}. 
If~\eqref{eq:opt_ref} is solved with a standard convex solver, the solver will most likely end in the same local minimum as the closed-loop MPC scheme. 
Thus, even if the convexity condition is not satisfied, the proposed scheme~\eqref{eq:limon} is still a good choice.   
\end{rem}
%
%!TEX root = ./PeriodicTracking_Automatica.tex
%%%%%%%%%%%%%%%%%%%%%%%%%%%%%%%%%%%%%%%%%%%%%%%%%%%%%%%%%%%%%%%%%%%%%%%%%%%%%%% 
%
The following theorem establishes \textit{exponential} stability of the optimal \textit{reachable} trajectory $x^{T*}$ given suitable terminal ingredients (Ass.~\ref{ass:term_gen}) and the convexity condition on the set of feasible periodic orbits (Ass.~\ref{ass:unique_convex}), which is the main result of this paper. 
This result generalizes and unifies the results in~\cite{limon2018nonlinear,limon2016mpc}, by considering \textit{nonlinear} dynamics, \textit{periodic} reference trajectories, establishing \textit{exponential} stability, and unifying the consideration of different terminal ingredients (Ass.~\ref{ass:term_gen}, Lemma~\ref{lemma:LPV_term}, Prop.~\ref{prop:TEC}). 
\begin{thm}
\label{thm:track}
Let Assumptions~\ref{ass:term_gen} and \ref{ass:unique_convex} hold. 
Assume that $h$ is bounded on $\mathcal{Z}_r$ and that the Problem~\eqref{eq:limon} is feasible at $t=0$. 
Then the Problem~\eqref{eq:limon} is recursively feasible for the resulting closed-loop system~\eqref{eq:close_limon}, for arbitrary target signals $y^e$.
Furthermore, for a $T$-periodic target signal $y^e$,  the optimal reachable trajectory $x^{T*}$ is (uniformly) exponentially stable for the resulting closed-loop system~\eqref{eq:close_limon}. 
\end{thm}
\begin{pf}
\textbf{Part I:} Recursive Feasibility: 
It suffices to note that feasibility of~\eqref{eq:limon} does not depend on the target signal $y^e$. 
Correspondingly, the input sequence $u_{\cdot|t+1}$ in Theorem~\ref{thm:MPC} with the shifted reference $r_{k|t+1}=r^*_{k+1|t}$ is a feasible solution to~\eqref{eq:limon}. \\
\textbf{Part II:}  Stability: 
Consider a periodic target signal $y^e_{k|t}=y^e_{k-1|t+1}$, which is denoted by $y^e_{t+k}:=y^e_{k|t}$. 
Thus, the minimizer of~\eqref{eq:opt_ref} is a periodic trajectory, i.e., $x^{T*}_{k+t}:=x^{T*}_{k|t}=x^{T*}_{k-1|t+1}$, and we write $J_T(r,t):=J_T(r,y^e_{\cdot|t})$, $V_T:=V_T(y^e_{\cdot|t})$, with $J_T$ (periodically) time-varying in the second argument and $V_T$ constant in time. 
Define the candidate Lyapunov function
$W_t:=W_T(x_t,y^e_{\cdot|t})-V_T$
and the error $e^T_t:=x_t-x^{T*}_t$.  
In the following, we show that there exists a positive constant $\alpha_W$, such that  
\begin{subequations}
\label{eq:W}
\begin{align}
\label{eq:W_1}
&W_{t+1}\leq W_t-\|x_t-x^{r*}_{0|t}\|_Q^2,\\
\label{eq:W_2}
&\alpha_W \|e^T_t\|_Q^2\leq W_t\leq c_v\|e^T_t\|_Q^2,
\end{align}
\end{subequations}
holds for all $x_t$ such that Problem~\eqref{eq:limon} is feasible.
The shifted reference $r_{\cdot|k+1}$  in Part I  satisfies 
$J_T(r_{\cdot|t+1},t+1)=J_T(r^*_{\cdot|t},t)$. 
Thus, feasibility in combination with~\eqref{eq:V} implies
$W_{t+1}-W_t\leq  -\ell(x_t,u_t,r^*_{0|t})$,
which implies~\eqref{eq:W_1}.
Lipschitz continuity of $g$ implies  
\begin{align}
\label{eq:Lipschitz}
&\|x^{r*}_{0|t}-{x}^{T*}_{0|t}\|_Q\leq \|x^{r*}_{\cdot|t}-{x}^{T*}_{\cdot|t}\|_Q\\
=&\|g_x({y}^{r*}_{\cdot|t})-g_x(y^{T*}_{\cdot|t})\|_Q\leq L_g\|{y}^{r*}_{\cdot|t}-y^{T*}_{\cdot|t}\|_S, \nonumber
\end{align}
with some constant $L_g$. 
Thus, strict convexity (compare~\eqref{eq:unique}) implies 
\begin{align*}
&J_T(r^*_{\cdot|t},t)-V_T\geq \|y^{*T}_{\cdot|t}-y^{r*}_{\cdot|t}\|_S^2\\
\geq &1/{L_g^2}\|x^{T*}_{\cdot|t}-x^{r*}_{\cdot|t}\|_Q^2
\geq 1/{L_g^2}\|x^{T*}_{0|t}-x^{r*}_{0|t}\|_Q^2. 
\end{align*}
Correspondingly, using the fact that $a^2+b^2\geq 1/2(a+b)^2$ for all $a,b\in\mathbb{R}$ yields the lower bound
\begin{align*}
&W_t
\geq\|x_t-x^{r*}_{0|t}\|_Q^2+J_T(r^*_{\cdot|t},t)-V_T\\ 
\geq &\|x_t-x^{r*}_{0|t}\|_Q^2+1/L_g^2\|x^{*T}_{0|t}-x^{r*}_{0|t}\|_Q^2
\geq \alpha_W\|e^T_t\|_Q^2,
\end{align*}
with $\alpha_W=\frac{1}{2}\min\{1,1/L_g^2\}$. 
In case $\|e^T_t\|_Q^2\leq \epsilon^2$, Inequality~\eqref{eq:value_quad_bound} in Assumption~\ref{ass:term_gen} ensures that $r_{\cdot|t}=r^{T*}_{\cdot|t}$ is a feasible solution to~\eqref{eq:limon}, which satisfies
$W_t\leq  c_u\|e^T_t\|_Q^2$. 
As in Theorem~\ref{thm:MPC}, compact constraints together with this local upper bound imply the imply the upper bound in~\eqref{eq:W_2}, compare~\cite[Prop.~2.16]{rawlings2017model}.
%\cite[Prop.~2.18]{rawlings2012postface}.
Inequalities~\eqref{eq:W} imply (uniform) stability of $x^{T*}$ for the closed-loop system, but not necessarily asymptotic or exponential stability. \\
\textbf{Part III: }Exponential stability -  case distinction:\\
\textbf{Case 1:}
Consider 
\begin{align}
\label{eq:case_1}
\|x_t-x^{r*}_{0|t}\|_Q^2\geq \gamma\|y^{r*}_{\cdot|t}-y^{T*}_{\cdot|t}\|_S^2,
\end{align}
 with a later specified positive constant $\gamma$. 
Then, \eqref{eq:W_1} and \eqref{eq:Lipschitz} imply
\begin{align*}
&W_{t+1}-W_t\stackrel{\eqref{eq:W_1}}{\leq} -\|x_t-x^{r*}_{0|t}\|_Q^2\\
\stackrel{\eqref{eq:case_1}}{\leq}& -1/2(\|x_t-x^{r*}_{0|t}\|_Q^2+\gamma\|y^{r*}_{\cdot|t}-y^{T*}_{\cdot|t}\|_S^2)\\
\stackrel{\eqref{eq:Lipschitz}}{\leq}&-1/{2}(\|x_t-x^{r*}_{0|t}\|_Q^2+ {\gamma}/L_g^2\|x^{r*}_{0|t}-x^{T*}_{0|t}\|_Q^2)\\ 
\leq&- 1/{4}\min\left\{1, {\gamma}/{L_g^2}\right\}\|x_t-x^{T*}_t\|_Q^2. 
\end{align*}
\textbf{Case 2:} Assume  
\begin{align}
\label{eq:case_2}
\|x_t-x^{r*}_{0|t}\|_Q^2\leq  \gamma\|y^{r*}_{\cdot|t}-y^{T*}_{\cdot|t}\|_S^2. 
\end{align} 
Boundedness of $h$ on $\mathcal{Z}_r$  implies that there exists a constant $y_{\max}$, such that $\|y^r_{\cdot|t}-\tilde{y}^r_{\cdot|t}\|_S^2\leq y_{\max}$, for any trajectories $y^r$, $\tilde{y}^r$ that satisfy the constraints in~\eqref{eq:opt_ref}.  
This implies
\begin{align*}
\|x_t-x^{r*}_{0|t}\|_Q^2\stackrel{\eqref{eq:case_2}}{\leq} \gamma \|y^{r*}_{\cdot|t}-y^{T*}_{\cdot|t}\|_S^2\leq \gamma y_{\max}. 
\end{align*}
For $\gamma\leq \gamma_1:={\epsilon}^2/{y_{\max}}$, we have $\|x_t-x^{r*}_{0|t}\|_Q^2\leq \epsilon^2$. 
Thus, Assumption~\ref{ass:term_gen} implies 
\begin{align}
\label{eq:bound_x_next}
& \|x^*_{1|t}-x^{r*}_{1|t}\|_Q^2\leq 
J_N(x^*_{\cdot|t},u^*_{\cdot|t},r^*_{\cdot|t})\\
&\stackrel{\eqref{eq:value_quad_bound}}{\leq}  c_u\|x_t-x^{r*}_{0|t}\|_Q^2
\stackrel{\eqref{eq:case_2}}{\leq} \gamma c_u\|y^{r*}_{\cdot|t}-y^{T*}_{\cdot|t}\|_S^2\leq \gamma c_u y_{\max}. \nonumber
\end{align}
For $\gamma\leq \gamma_2:= \epsilon^2/(4c_uy_{\max})\leq \gamma_1$ this implies 
$\|x_{t+1}-{x}^{r*}_{1|t}\|_Q\leq  \epsilon/2$. 
 Consider $y^r_{k|t+1}=h(r_{k|t+1})$, where  $r_{\cdot|t+1}$ is the candidate reference from Part~I of the proof. 
At time $t+1$, define an auxiliary reference  
\begin{align}
\label{eq:y_r}
\hat{y}^r:=\beta y^r_{\cdot|t+1} +(1-\beta)y^{T*}_{\cdot|t+1},~ \beta\in[0,1],
\end{align}
with the corresponding state and input trajectory $\hat{x}^r=g_x(\hat{y}^r)$, $\hat{u}^r=g_u(\hat{y}^r)$, $\hat{r}=(\hat{x}^r,\hat{u}^r)$. 
Convexity (Ass.~\ref{ass:unique_convex}) ensures that the auxiliary reference $\hat{r}$ is a feasible solution to~\eqref{eq:opt_ref} at $t+1$. 
This definition implies
\begin{align}
\label{eq:1_beta}
\hat{y}^r-y^r_{\cdot|t+1}=(1-\beta)(y^{T*}_{\cdot|t+1}-y^r_{\cdot|t+1}).
\end{align}
The cost $J_T$ satisfies 
\begin{align}
\label{eq:decrease_J_T}
&J_T(\hat{r},t+1)- J_T(r^*_{\cdot|t},t)\nonumber\\
=&(\hat{y}^r-y^r_{\cdot|t+1})^\top S (\hat{y}^r+y^r_{\cdot|t+1}-2y^e_{\cdot|t+1})\nonumber\\
\stackrel{\eqref{eq:y_r}}{=}&(1-\beta)(y^{T*}_{\cdot|t+1}-y^r_{\cdot|t+1})^\top S\nonumber\\
&((1+\beta)y^r_{\cdot|t+1}+(1-\beta)y^{T*}_{\cdot|t+1}-2y^e_{\cdot|t+1})\nonumber\\
=&-(1-\beta^2)\|y^{r*}_{\cdot|t}-y^{T*}_{\cdot|t}\|_S^2\nonumber\\
&+(1-\beta)(y^{T*}_{\cdot|t+1}-y^r_{\cdot|t+1})  \nabla_{y^r}J_T(y^r,y^e)|_{y^r=y^{T*}}\nonumber\\
\stackrel{\eqref{eq:convex_grad}}{\leq}& -(1-\beta^2)\|y^{r*}_{\cdot|t}-y^{T*}_{\cdot|t}\|_S^2.
\end{align}
Lipschitz continuity (compare~\eqref{eq:Lipschitz}) implies
 \begin{align*}
&\|x_{t+1}-\hat{x}^r_0\|_Q
\leq \|x_{t+1}-x^{r*}_{1|t}\|_Q+\|x^{r*}_{1|t}-\hat{x}^r_0\|_Q\\
\leq & \epsilon/2+L_g\|y^r_{\cdot|t+1}-\hat{y}^r\|_S\\
\stackrel{\eqref{eq:1_beta}}{=} & \epsilon/2 + L_g(1-\beta)\|y^r_{\cdot|t+1}-y^{T*}_{\cdot|t+1}\|_S.
\end{align*}
For $\beta\in[ {\beta}_1,1]$ with $\beta_1:=1-{\epsilon}/({2L_g\sqrt{y_{\max}}})$, this implies 
$\|x_{t+1}-\hat{x}^r_0\|_Q\leq \epsilon$. 
Thus, Assumption~\ref{ass:term_gen} ensures that there exists some state and input sequence $(\hat{x},\hat{u})$, such that  $(\hat{x},\hat{u},\hat{r})$ is a feasible solution to~\eqref{eq:limon} at time $t+1$ and the tracking cost satisfies
\begin{align}
\label{eq:decrease_J_N}
&J_N(\hat{x},\hat{u},\hat{r})
\stackrel{\eqref{eq:value_quad_bound}}{\leq} c_u\|x_{t+1}-\hat{x}^r_0\|_Q^2\\
\leq& 2c_u(\|x_{t+1}-x^{r*}_{1|t}\|_Q^2+\|x^{r*}_{1|t}-\hat{x}^r_0\|_Q^2)\nonumber\\
{\leq}& 2c_u\|x_{t+1}-x^{r*}_{1|t}\|_Q^2+2c_uL_g^2\|y^r_{\cdot|t+1}-\hat{y}^r\|_S^2\nonumber\\
\stackrel{\eqref{eq:1_beta}}{\leq}& 2c_u\|x_{t+1}-x^{r*}_{1|t}\|_Q^2
+2c_uL_g^2(1-\beta)^2\|y^{r*}_{\cdot|t}-y^{T*}_{\cdot|t}\|_S^2,\nonumber
\end{align}
where the second to last inequality follows from Lipschitz continuity, compare~\eqref{eq:Lipschitz}. 
Correspondingly, we have
\begin{align*}
&W_{t+1}-W_t\\
&\leq J_N(\hat{x},\hat{u},\hat{r})+J_T(\hat{r},t+1)
-J_T({r}^*_{\cdot|t},t)-\|x_t-x^{r*}_{0|t}\|_Q^2\\
&\stackrel{\eqref{eq:decrease_J_T},\eqref{eq:decrease_J_N}}{\leq} 2c_u\|x_{t+1}-x^{r*}_{1|t}\|_Q^2-\|x_t-x^{r*}_{0|t}\|_Q^2\\
&-\underbrace{((1-\beta^2)-2c_uL_g^2(1-\beta)^2)}_{=:c_2(\beta)}\|y^{r*}_{\cdot|t}-y^{T*}_{\cdot|t}\|_S^2\\
&\stackrel{\eqref{eq:Lipschitz},\eqref{eq:bound_x_next}}{\leq}  (2c_u^2\gamma
-c_2(\beta)/2)\|y^{r*}_{\cdot|t}-y^{T*}_{\cdot|t}\|_S^2\\
&-c_2(\beta)/(2L_g^2)\|x^{r*}_{0|t}-x^{T*}_{0|t}\|_Q^2-\|x_t-x^{r*}_{0|t}\|_Q^2\\
&{\leq}(2c_u^2\gamma-c_2({\beta})/2)\|y^{r*}_{\cdot|t}-y^{T*}_{\cdot|t}\|_S^2\\
&-\min\{1/2,c_2(\beta)/(4L_g^2)\}\|x_t-x^{T*}_{0|t}\|_Q^2.
\end{align*}	
Let $\beta={\beta}_2:=\arg\max_{\beta\in[{\beta}_1,1]}c_2(\beta)$, with $c_2({\beta}_2)>0$. 
For $\gamma\leq \gamma_3:= {c_2({\beta}_2)}/(4c_u^2)$, this implies 
\begin{align*}
&W_{t+1}-W_t
\leq -\min\left\{1/2,c_2({\beta}_2)/(4L_g^2)\right\}\|x_t-x^{T*}_t\|_Q^2.
\end{align*}
\textbf{Combine:} Combining these two cases yields
\begin{align}
\label{eq:W_4}
W_{t+1}\leq& W_t-\gamma_T\|x_t-x^{T*}_t\|_Q^2,\\
\gamma_T:=&\min\left\{\frac{c_2({\beta}_2)}{4L_g^2},\frac{1}{4},\frac{\gamma}{4L_g^2}\right\},~\gamma:=\min\{\gamma_1,\gamma_2,\gamma_3\}.\nonumber
\end{align}
Uniform exponential stability follows using Inequalities~\eqref{eq:W_2},  \eqref{eq:W_4} and Lyapunov arguments. 
$\hfill\square$
\end{pf}
Theorem~\ref{thm:track} ensures exponential stability of  the optimal reachable trajectory $x^{T*}$ by showing quadratic lower and upper bounds and an exponential decay of the Lyapunov function $W_t:=W_T(x_t,y^e_{\cdot|t})-V_T$. 
The exponential decay of $W_t$  is shown by utilizing two distinct candidate solutions, namely $(\hat{x},\hat{u},\hat{r})$ and the standard candidate solution from Theorem~\ref{thm:MPC}.
In particular, we distinguish whether the tracking error $\|x^{r*}_{0|t}-x_t\|_Q^2$ is large/small ($\gamma$) compared to the output tracking cost $J_T=\|y^{r*}_{\cdot|t}-y^e_{\cdot|t}\|_S^2$.
If the reference tracking error is large, then the standard candidate solution, e.g. used in Theorem~\ref{thm:MPC}, ensures a sufficient exponential decrease in the Lyapunov function $W_t$. 
On the other hand, if the reference tracking error is small enough ($\gamma y_{\max}$), then the convexity condition (Ass.~\ref{ass:unique_convex}) ensures that the artificial reference $r$ can be incrementally  ($\beta$) moved towards the optimal reachable reference $r^{T*}$, which decreases the output tracking cost $J_T$. 
The local quadratic bound~\eqref{eq:value_quad_bound} (Ass.~\ref{ass:term_gen}) on the value function $V$ ensures that the optimization problem is feasible with the incrementally moved reference $\hat{r}$ and that the increase in the tracking cost $J_N$ is quadratically bounded. 
Finally, there exists a sufficiently small change ($\beta_2$), such that this auxiliary candidate solution $(\hat{x},\hat{u},\hat{r})$ ensures an exponential decay in $W_t$.

\begin{rem}
\label{rk:robust}
Similar to the derivations in~\cite{limon2018nonlinear,limon2016mpc}, Theorem~\ref{thm:track} assumes no model mismatch, which is rarely the case in practical applications. 
To ensure \textit{robust} recursive feasibility despite disturbances, the MPC problem~\eqref{eq:limon} needs to be adjusted using constraint tightening techniques from robust MPC. 
A corresponding formulation for nonlinear robust tracking MPC systems can be found in the recent paper~\cite{nubert2020safe}, which combines the formulation presented in Section~\ref{sec:limon_alpha} with the nonlinear robust MPC formulation in~\cite{Robust_TAC_19}. 

In addition to possible feasibility issues, model mismatch typically also implies non-zero offset, even in case of constant references. 
For the special case of setpoint tracking ($T=1$), this issue is typically resolved using \textit{offset-free} MPC formulations (cf.~\cite{morari2012nonlinear} and references therein), which rely on a disturbance estimator for constant offsets.
In order to transfer this concept to $T$-periodic trajectories, the dimension of the disturbance model must be correspondingly increased. 
An alternative approach to ensure offset-free tracking is to use a parameter estimation scheme with an adaptive MPC formulation, under appropriate assumptions on the model mismatch, compare e.g. \cite{bujarbaruah2019adaptive}.  
Extending the proposed approach to ensure offset free tracking despite deterministic model mismatch is an open problem. 
\end{rem}
\begin{rem}
\label{rk:ref_govener}
The proposed approach can be extended to stabilize the \textit{economically} optimal reachable reference $r$, as an extension to the linear approach in~\cite{limon2014single}, assuming that \eqref{eq:opt_ref} remains (strictly) convex (Ass.~\ref{ass:unique_convex}). 
The extension of the approaches in~\cite{muller2013economic,fagiano2013generalized} for \textit{economically} optimal steady-state operation, to optimal dynamic operation based on this result is part of current research.  
Recently, in~\cite{berberich2020track} for the special case of linear systems, the analysis in Theorem~\ref{thm:track} was extended to stage costs $\ell$ with $Q$ semi-definite using suitable observability conditions, which is especially relevant for input-output models. 
 
The consideration of nonperiodic dynamic target signals in this framework is still an open topic.
Setpoint stabilization is a special case in Theorem~\ref{thm:track} with $T=1$,  $\mathcal{R}(r)=r$ and the feasible steady-state manifold $\mathcal{Z}_r$, which has also been considered in~\cite{limon2018nonlinear}. 
Compared to~\cite{limon2018nonlinear}, Theorem~\ref{thm:track} ensures exponential stability with convergence rate $\frac{c_v-\gamma_T}{c_v}<1$ and Lyapunov function $W_t$, while~\cite{limon2018nonlinear} uses a proof of contradiction to establish convergence. 
Furthermore, the general assumptions on the terminal ingredients (Ass.~\ref{ass:term_gen}) allow us to use the continuously parameterized and thus differentiable terminal cost $V_f$  based on Lemma~\ref{lemma:LPV_term} (cf.~\cite{JK_QINF}), instead of  partitioning $\mathcal{Z}_r$, as done in~\cite{limon2018nonlinear}.    
The practicality of this result is demonstrated in the numerical examples in Section~\ref{sec:num}. 
In the setpoint stabilization case $(T=1)$ with $N=0$, the candidate solutions in Theorem~\ref{thm:track} correspond to an inner-loop controller  $k_f(x,r)$ with a corresponding (explicit) reference governor (cf.~\cite{garone2017reference}) for $r$ given by~\eqref{eq:y_r}, compare the numerical example in Section~\ref{sec:num}. 
\end{rem}
\begin{rem}
\label{rk:terminal_zero}
We note that this result guarantees exponential stability of the optimal reachability trajectory $x^{T*}$ using either terminal equality constraints (Prop.~\ref{prop:TEC}) or a terminal cost (Lemma~\ref{lemma:LPV_term}), respectively. 
However, the quantitative bounds regarding convergence rate may differ significantly. 
In particular,  including suitably designed terminal ingredients greatly improves the closed-loop performance,  
which is demonstrated in the numerical examples in Section~\ref{sec:num}, compare also examples in~\cite{JK_QINF}. 
\end{rem}

%!TEX root = ./PeriodicTracking_Automatica.tex
%%%%%%%%%%%%%%%%%%%%%%%%%%%%%%%%%%%%%%%%%%%%%%%%%%%%%%%%%%%%%%%%%%%%%%%%%%%%%%%
\subsection{Online optimized terminal set size}
\label{sec:limon_alpha}
In case we use a terminal cost and terminal set (Lemma~\ref{lemma:LPV_term}), the fact that $\mathcal{Z}_r$ is chosen in advance  (as also done in~\cite{limon2018nonlinear,limon2016mpc}) can be disadvantageous. 
In particular, similar to setpoint stabilization in~\cite{chen1998quasi}, the terminal set size $\alpha$ in Lemma~\ref{lemma:LPV_term} is  the minimum of two values: $\alpha_1$ and $\alpha_2$.  
The first ($\alpha_1$) is independent of $\mathcal{Z}_r$ and needs to be such that~\eqref{eq:term_dec} holds (cf.~\cite[Alg.~1]{JK_QINF}). 
The second ($\alpha_2$) ensures constraint satisfaction~\eqref{eq:term_con} and depends on the difference between $\mathcal{Z}$ and $\mathcal{Z}_r$. 
Thus, by choosing $\mathcal{Z}_r$ we trade achievable terminal set size ($\alpha_2$) (and hence convergence speed of the closed-loop state) against operation close to the boundary of the constraint set $\mathcal{Z}$. 
In the following, we show how $\alpha$ can be optimized online, instead of using a preassigned reference constraint set $\mathcal{Z}_r$. 
In~\cite{simon2014reference}, a similar dynamic scaling of the terminal set has been suggested for linear setpoint tracking with a polytopic terminal set (albeit with a different motivation). 
The proposed formulation is geared towards the terminal ingredients from Lemma~\ref{lemma:LPV_term} and polytopic constraints of the form $\mathcal{Z}=\{r\in\mathbb{R}^{n+m}|~L_i r\leq l_i,~i=1,\dots,n_z\}$. 
To this end, we define the following functions
\begin{align*}
L_{PK,i}(r):=\|P_f^{-1/2}(r)[I_n,K_f^\top(r)]L_{i}^\top \|,~i=1,\dots,n_z. 
\end{align*}
The modified optimization problem is given by
\begin{subequations}
\label{eq:limon_alpha}
\begin{align}
&\min_{{u_{\cdot|t},r_{\cdot|t},\alpha^s_t}}J_N(x_{\cdot|t},u_{\cdot|t},r_{\cdot|t})+J_T(r_{\cdot|t},y^e_{\cdot|t})\nonumber\\
\text{s.t. }&x_{k+1|t}=f(x_{k|t},u_{k|t}),~
 x_{0|t}=x_t,\\
 &(x_{k|t},u_{k|t})\in\mathcal{Z},~k=0,\dots,N-1~\\
\label{eq:limon_alpha_term}
& V_f(x_{N|t},r_{N|t})\leq (\alpha_t^s)^2,~
%\label{eq:limon_alpha_s}
\sqrt{\alpha_{\min}}\leq \alpha^s_t\leq \sqrt{\alpha_1},\\
\label{eq:limon_alpha_2}
&x^r_{j+1|t}=f(x^r_{j|t},u^r_{j|t}),~r_{l+T|t}=r_{l|t},  \\
\label{eq:limon_alpha_1}
&L_{i}r_{j|t}+L_{PK,i}(r_{j|t})\alpha^s_t\leq l_i,~i=1,\dots,n_z\\
& j=0,\dots,T-1,~ l=0,\dots,\max\{0,N-T\},.\nonumber
\end{align}
\end{subequations}
Compared to~\eqref{eq:limon}, the reference constraints~\eqref{eq:limon_Zr} and in particular the polytopic constraints $\mathcal{Z}_r$ are replaced by~\eqref{eq:limon_alpha_2} and~\eqref{eq:limon_alpha_1} and we have one additional optimization variable $\alpha_s=\sqrt{\alpha}$, where $\alpha_{\min}>0$  is needed to avoid robustness issues and retain the closed-loop properties of the original scheme (Thm.~\ref{thm:track}). 
The expression $P_f^{-1/2}(r)$ in~$L_{PK,i}$ denotes any matrix square root of $X(r)=P_f^{-1}(r)$, which can be computed using the command \textit{sqrtm}$(X)$ or $\textit{chol}(X)$ in  Matlab. 
In the numerical examples (Sec.~\ref{sec:num}), we implement~\eqref{eq:limon_alpha_1} using the symbolic Cholesky decomposition of $X(r)=X_0+\sum_i\theta_i(r)X_i$, which is suitable for automatic differentiation used in CasADi~\cite{andersson2019casadi}.  %
\begin{prop}
\label{prop:limon_alpha}
Suppose there exist matrices $P_f$, $K_f$ and a constant $\alpha>0$, such that   condition~\eqref{eq:term_dec} holds for all
$r,r^+\in\mathcal{Z}$, $V_f(x,r)\leq \alpha_1$ with  $x^{r+}=f(x^r,u^r)$, $x^+=f(x,u^r+K_f(r)(x-x^r))$.
Assume further that  $h$ is bounded on $\mathcal{Z}$ and Ass.~\ref{ass:unique_convex} holds with $\mathcal{Z}_r$ in~\eqref{eq:opt_ref} replaced by %
\begin{align*}
\tilde{\mathcal{Z}}_r=\{r|~L_ir+L_{PK,i}(r)\sqrt{\alpha_{\min}}\leq l_i,~i=1,\dots,n_z\}.
\end{align*}
Then the MPC scheme based on~\eqref{eq:limon_alpha} satisfies the theoretical properties in Theorem~\ref{thm:track}. 
Consider the scheme~\eqref{eq:limon} with a given constraint set $\mathcal{Z}_r$ and correspondingly computed $\alpha_2$. 
If $\alpha_2\in(\alpha_{\min},\alpha_1)$, the scheme~\eqref{eq:limon_alpha} can stabilize references $r\notin\mathcal{Z}_r$ and has a larger region of attraction.  
\end{prop}
\begin{pf}
First, note that~\eqref{eq:limon_alpha_1} is equivalent to~\eqref{eq:term_con}, compare e.g.~\cite[Equation~(10)]{conte2016distributed} based on the support function. 
Thus, every feasible solution to~\eqref{eq:limon} is also a feasible solution to~\eqref{eq:limon_alpha} with $\alpha^s_t=\sqrt{\alpha}$.   
Recursive feasibility follows with the same candidate input by using $\alpha^s_{t+1}=\alpha^{s*}_t$, where $\alpha^{s*}_t$ denotes the solution to~\eqref{eq:limon_alpha}. 
Parts II and III of Theorem~\ref{thm:track} remain true since Inequality~\eqref{eq:value_quad_bound} holds for all references $r_{\cdot|t}\in\tilde{\mathcal{Z}}_r$ with $\epsilon=\sqrt{\alpha_{\min}/c_u}>0$. 
Satisfaction of Ass.~\ref{ass:unique_convex} with $\tilde{\mathcal{Z}}_r$ as defined above ensures that the reference $\hat{y}^r$~\eqref{eq:y_r} satisfies the constraints in~\eqref{eq:limon_alpha} with $\alpha^s=\sqrt{\alpha_{\min}}$. 
Furthermore, $\alpha^s>\sqrt{\alpha_2}$ provides a larger terminal set and thus enlarges the set of feasible initial conditions. 
For $\alpha^s<\sqrt{\alpha_2}$ we can consider references $r\notin\mathcal{Z}_r$ (close to the boundary of $\mathcal{Z}$) and thus provide a feasible solution for initial conditions close to the constraints and stabilize references $r\notin\mathcal{Z}_r$. 
$\hfill\square$
\end{pf}
Proposition~\ref{prop:limon_alpha} essentially confirms that this modification preserves the theoretical properties in Theorem~\ref{thm:track}.
In summary, the optimization over $\alpha^s$ provides an additional degree of freedom which can significantly enlarge the terminal set and lead to faster convergence.
For numerical reasons, one can also replace the constraint~\eqref{eq:limon_alpha_1} with the more conservative constraint
\begin{align}
\label{eq:l_max}
&L_{\max,i}\alpha^s_t+L_{i}r_{j|t}\leq l_i,~ j=0,\dots T-1,\\
&L_{\max,i}:=\max_{r\in\mathcal{Z}}L_{PK,i}(r), ~i=1,\dots,n_z. \nonumber
\end{align}
This formulation also retains the properties in Theorem~\ref{thm:track} and the constraint~\eqref{eq:l_max} is linear in the optimization variables $\alpha^s,~r$.  
This constraint on the reference $r$ is similar to the constraint tightening in robust MPC with a variable tube size $\alpha^s$.  
In the linear polytopic setpoint tracking case, a polyhedral invariant set for tracking  can be used to constrain the reference $r$ and the terminal state $x$, compare~\cite[Ass.~2]{limon2008mpc}. 
In~\cite{simon2014reference} the complexity of the polytopic characterization is reduced by online optimizing a scaling of the polytopic terminal set. 
The proposed approach, especially the simplified formula~\eqref{eq:l_max}, can be viewed as an extension of this approach to nonlinear systems with ellipsoidal terminal sets. %
For comparison, in~\cite{limon2018nonlinear} the reference constraint set $\mathcal{Z}_r$ is partitioned and a fixed size $\alpha_i$ is considered for each partition $\mathcal{Y}_i$, which is conservative compared to the online optimized value $\alpha^s$, compare Fig.~\ref{fig:CSTR_setpoint_alpha} in Section~\ref{sec:num}.   
{\begin{rem}
\label{rk:linear_onlinealpha}
The proposed approach with online optimization of $\alpha^s$ also provides a simple means to avoid terminal equality constraints as a valuable extension of the linear periodic tracking MPC in~\cite{limon2016mpc}. 
In case of linear system dynamics with a polytopic terminal set and online optimized $\alpha^s$ (cf.~\cite{simon2014reference}), the overall optimization problem is a quadratic program (QP) with one additional scalar variable $\alpha^s$ and linear inequality constraints instead of the terminal equality constraint. 
\end{rem}
\begin{rem}
\label{rk:alpha_rho}
It is possible to further relax the tightened reference constraints~\eqref{eq:limon_alpha_1} or \eqref{eq:l_max}, by taking into account the fact that the terminal set is contractive with some constant $\rho\in(0,1)$, i.e.,
\begin{align}
\label{eq:contract}
&V_f(f(x,k_f(x,r)),r^+)\leq \rho^2V_f(x,r),\\
&\forall ~V_f(x,r)\leq \alpha_1,~r,r^+\in\mathcal{Z}, x^{r+}=f(x^r,u^r),\nonumber
\end{align}
 compare~\cite[Prop. 1]{JK_QINF}. 
In particular, by redefining $\alpha^s=\sqrt{\alpha}-\sqrt{\alpha_{\min}}$ we can replace the constraints~\eqref{eq:limon_alpha_term},\eqref{eq:limon_alpha_1} by the following constraints
\begin{subequations}
\label{eq:limon_alpha_rho}
\begin{align*}
&V_f(x_{N|t},r_{N|t})\leq (\alpha^s_t+\sqrt{\alpha_{\min}})^2,\\
&L_{\max,i}(\alpha^s_t \rho^{\text{mod}(j+T-N, T)}+\sqrt{\alpha_{\min}})\leq l_i-L_{i}r_{j|t}, \\
&\alpha^s_t\in[0,\sqrt{\alpha_1}-\sqrt{\alpha}_{\min}],
\end{align*}
\end{subequations}
where $\text{mod}$ denotes  the modulo operator. 
The theoretical properties in Theorem~\ref{thm:track} and Proposition~\ref{prop:limon_alpha} remain valid with the candidate solution $\alpha^s_{t+1}=\rho\alpha^{s*}_t$, which satisfies $\alpha_{t+1}\geq \max\{\rho^2\alpha^*_t,\alpha_{\min}\}$, with $\alpha^*_t=(\alpha^{s*}_t+\sqrt{\alpha}_{\min})^2$. 
These relaxed constraints are especially useful in transient operation with active constraints on the reference $r$ and $\rho^T\ll 1$.
\end{rem}

%!TEX root = ./PeriodicTracking_Automatica.tex
%%%%%%%%%%%%%%%%%%%%%%%%%%%%%%%%%%%%%%%%%%%%%%%%%%%%%%%%%%%%%%%%%%%%%%%%%%%%%%%
%\subsection{Asynchronous reference update} % $\alpha$}
%\sub
\subsection{Partially decoupled reference updates}  
\label{sec:partial}
In the following, we demonstrate that the joint stabilization and trajectory planning (Sec.~\ref{sec:ext}) can be partially decoupled, which can significantly reduce the online computational demand. \\
\textit{Motivation: }
The main premise of the proposed approach (Sec.~\ref{sec:ext})  is that the operating conditions change on a time scale similar to the system dynamics, which in turn necessitates online updates of the reference trajectory. 
The most challenging problems are those, where the operating conditions change at a similar time scale to the system dynamics, while the target signal and hence the optimal system operation is determined based on long term considerations that involve a significantly larger time scale, i.e., the period length $T$ is very large.
An example of such a multi time-scale problem would be energy systems, compare e.g.~\cite{kumar2018stochastic}, where real time decisions are made every $5~min$, while the planning horizon is 7 days yielding $T\geq 2\cdot 10^3$. 
For such problems, it is vital that the reference $r$ is updated frequently, while at the same time it may be computationally too expensive to solve the joint planning and regulation problem~\eqref{eq:limon} in each time step $t$. \\
\textit{Continuity terminal cost: }
In the following, we demonstrate how the optimization problem~\eqref{eq:limon} can be decomposed into two partially decoupled optimization problems that may be solved at different time scales by using the following continuity property of the terminal cost.
\begin{prop}
\label{prop:term_cost_cont}
Suppose that the conditions in Lemma~\ref{lemma:LPV_term} are satisfied with $\mathcal{Z}_r=\mathcal{Z}$, $P_f=X^{-1}$, $K_f=YP_f$,
\begin{align*}
X(r)=X_0+\sum_{j=1}^p\theta_j(r)X_j,~Y(r)=Y_0+\sum_{j=1}^p\theta_j(r)Y_j,
\end{align*}
with $\theta_j$  continuously differentiable.  
Then there exists a constant $L_p$, such that for any $r,~\tilde{r}\in\mathcal{Z}$ the terminal cost $V_f$ satisfies the following continuity condition
\begin{align}
\label{eq:term_cont}
&\sqrt{V_f(x,\tilde{r})}\\
\leq& \sqrt{V_f(x,r)}(1+L_p\|r-\tilde{r}\|)+\|x^r-\tilde{x}^r\|_{P_f(\tilde{r})}.\nonumber
\end{align}
\end{prop}
\begin{pf}
The fact that $\theta_j$ is  continuously differentiable directly implies that $X$ is continuously differentiable w.r.t. $r$.  
Thus, also the matrix $P_f=X^{-1}$ is  continuously differentiable in $r$ for any $r\in\mathcal{Z}$, using the fact $X$ is positive definite with uniform lower and upper bounds. 
This property in combination with compact constraints and uniform bounds on $P_f$ ensures that there exists a local Lipschitz constant $L_p\geq 0$, such that
\begin{align*}
P_f(\tilde{r})-P_f(r)\leq L_p P_f({r}) \|r-\tilde{r}\|,
\end{align*}
which implies
\begin{align*}
&\|x\|_{P_f(\tilde{r})}\leq \sqrt{\|x\|_{P_f(r)}^2+L_p\|x\|_{P_f(r)}^2\|r-\tilde{r}\|}\\
= &\|x\|_{P_f(r)}\sqrt{1+L_p\|r-\tilde{r}\|}
\leq  \|x\|_{P_f(r)}(1+L_p\|r-\tilde{r}\|),
\end{align*}
for any $r,~\tilde{r}\in\mathcal{Z}$. 
Thus, condition~\eqref{eq:term_cont} follows from
\begin{align*}
&\sqrt{V_f(x,\tilde{r})}=\|x-\tilde{x}^r\|_{P_f(\tilde{r})}\\
\leq &\|x-x^r\|_{P_f(\tilde{r})}+\|x^r-\tilde{x}^r\|_{P_f(\tilde{r})}\\
\leq &\sqrt{V_f(x,r)}(1+L_p\|r-\tilde{r}\|)+\|x^r-\tilde{x}^r\|_{P_f(\tilde{r})}. \quad\quad& \hfill\square
\end{align*}
\end{pf}
The result essentially follows from the continuously differentiable parametrization and the triangular inequality. 
In the special case of constant matrices $P_f$, condition~\eqref{eq:term_cont} is satisfied with $L_p=0$. 
In addition to the continuity property~\eqref{eq:term_cont}, we use the fact that the terminal cost is contractive with some factor $\rho\in(0,1)$, compare~\eqref{eq:contract}.

In the following, we summarize the basic approach to partially decouple the optimization problem~\eqref{eq:limon_alpha}. 
Suppose at time $t_i$, we have trajectories $x_{\cdot|t_i}$, $u_{\cdot|t_i}$, $r_{\cdot|t_i}$, $\alpha^s_{t_i}$, that satisfy the constraints in~\eqref{eq:limon_alpha}. 
For the next $M\in\mathbb{N}$ time-steps $t=t_i,\dots,t_i+M-1$, the tracking MPC considers the shifted reference $r^*_{\cdot|t_i}$, i.e.,
$r_{k|t}=r^*_{\text{mod}(k+t-t_i , T)|t_i}$ and the following updated terminal set size 
\begin{align}
\label{eq:alpha_s_update}
\alpha^{tr}_t=&\rho^{2(t-t_i)}\max\{{\alpha_{\min}},V_f(x_{N|t_{i}},r^*_{N|t_i})\},
\end{align}
 with the contraction rate $\rho$ according to~\eqref{eq:contract}. 
The closed-loop input is computed based on the following reference tracking MPC (similar to~\eqref{eq:MPC})
\begin{subequations}
\label{eq:MPC_M}
\begin{align}
&\min_{u_{\cdot|t}}J_N(x_{\cdot|t},u_{\cdot|t},r_{\cdot|t})\\
\text{s.t. }&x_{k+1|t}=f(x_{k|t},u_{k|t}),~(x_{k|t},u_{k|t})\in\mathcal{Z},\\
\label{eq:contract_term}
&V_f(x_{N|t},r_{N|t})\leq \alpha^{tr}_t,~x_{0|t}=x_t,\\
&k=0,\dots,N-1.\nonumber
\end{align}
\end{subequations}
Note that the contractive terminal constraint~\eqref{eq:contract_term} with $\alpha_t^{tr}$ according to~\eqref{eq:alpha_s_update} is similar to a contractive MPC~\cite{de2000contractive}.
In parallel,  the following reference optimization problem is solved at time $t_i$ in order to obtain an updated reference at $t_{i+1}=t_i+M$
\begin{subequations}
\label{eq:limon_alpha_M}
\begin{align}
&\min_{{r_{\cdot|t_{i+1}},\alpha^s_{t_{i+1}}}}J_T(r_{\cdot|t_{i+1}},y^e_{\cdot|t_{i+1}})\nonumber\\
\text{s.t. }&
\label{eq:alpha_s_M}
\rho^M\sqrt{\alpha^{tr}_{t_{i}}}(1+L_p\|r^*_{N+M|t_{i}}-r_{N|t_{i+1}}\|)\nonumber\\
&+\sqrt{V_f(x^{r*}_{N+M|t_{i}},r_{N|t_{i+1}})}
\leq  \alpha^s_{t_{i+1}},\\
\label{eq:limon_alpha_M_r}
&x^r_{j+1|t_{i+1}}=f(x^r_{j|t_{i+1}},u^r_{j|t_{i+1}}),\\
&r_{l+T|t_{i+1}}=r_{l|t_{i+1}},\\
\label{eq:limon_alpha_1_M}
&L_{i}r_{j|t_{i+1}}+L_{PK,i}(r_{j|t_{i+1}})\alpha^s_{t_{i+1}}\leq l_i,\\
\label{eq:limon_alpha_bounds_M}
&\sqrt{\alpha_{\min}}\leq \alpha^s_{t_{i+1}}\leq \sqrt{\alpha_1},\\
&i=1,\dots,n_z,~ j=0,\dots,T-1,\nonumber\\
&l=0,\dots,\max\{0,N+M-T\}.\nonumber
\end{align}
\end{subequations}
Note, that the constraints on the reference $r$ can be further relaxed using the formula in Remark~\ref{rk:alpha_rho}. 
Since we start to solve~\eqref{eq:limon_alpha_M} at time $t_i$, the target signal $y^e_{\cdot|t_{i+1}}$ is not yet available and instead the currently available target signal $y^e_{\cdot|t_i}$ needs to be shifted by $M$ time steps (assuming it is $T$-periodic). 
The overall algorithm is summarized in Alg.~\ref{alg:asyn}. 

\begin{algorithm}[h]
\caption{Partially decoupled reference updates}
\label{alg:asyn}
Execute at each time step $t_{i}=i\cdot M$, $i\in\mathbb{N}$
\begin{algorithmic}
\State Obtain $r^*_{\cdot|t_i}$ from reference planner~\eqref{eq:limon_alpha_M}.
\State Get $x_{N|t_i}$ from Tracking MPC~\eqref{eq:MPC_M}.
\State Compute $\alpha^{tr}_{t_i}$ using~\eqref{eq:alpha_s_update}.
\begin{algorithmic}
\State \textbf{Tracking MPC}
\For {$t=t_i, \hdots, t_i+M-1$}
\State Update $r_{\cdot|t}$, $\alpha^{tr}_t$.
\State Solve tracking MPC~\eqref{eq:MPC_M}.
\State Apply control input $u_t=u^*_{0|t}$.
\EndFor
\end{algorithmic}
\begin{algorithmic}
\State \textbf{Reference planner}
\State Obtain target signal $y^e_{\cdot|t_{i+1}}$.
\State Solve trajectory planning problem~\eqref{eq:limon_alpha_M}. 
\end{algorithmic}
\end{algorithmic}
%$i=i+1$
\end{algorithm}
The optimization problem~\eqref{eq:MPC_M} represents a standard tracking MPC (Sec.~\ref{sec:MPC}) that is executed in each time step $t$ with a fixed (periodic) reference trajectory $r$ and a shrinking terminal set. 
On the other hand, the optimization problem~\eqref{eq:limon_alpha_M} can be solved in the interval $[t_i,t_{i+M}]$, thus allowing to solve larger planning problems ($T>>1$) by updating the reference $r$ less frequently ($M\geq 1$). 
Condition~\eqref{eq:alpha_s_M} constrains how the updated reference $r$ may deviate from the previous solution, which partially couples the planning~\eqref{eq:limon_alpha_M} and regulation problem \eqref{eq:MPC_M}. 
Compared to a joint optimization, as in \eqref{eq:limon}, the practical convergence under changing operation conditions may be slower, as the reference is updated less frequently and the constraint~\eqref{eq:alpha_s_M} limits the rate of change, compare the numerical example in Sec.~\ref{sec:num}. 
However, the  partially decoupled updates in Alg.~\ref{alg:asyn} can significantly reduce the computational demand, especially in case of longer planning horizons $T$.  
\begin{prop}
\label{prop:asynch}
Suppose the conditions in Prop.~\ref{prop:limon_alpha} and~\ref{prop:term_cost_cont} hold and Alg.~\ref{alg:asyn} is initialized with $r^*_{\cdot|t_0},\alpha^{s*}_{t_0}$ satisfying~\eqref{eq:limon_alpha_M_r}--\eqref{eq:limon_alpha_bounds_M}, $\alpha^{tr}_{t_0}\leq \alpha^{s*}_{t_0}$ and $x_{t_0}$ such that \eqref{eq:MPC_M} is feasible.   
Then Alg.~\ref{alg:asyn} is recursively feasible for the resulting closed-loop system. 
Assume further\footnote{%
This condition excludes the special case where $L_{i,x}=-L_{i,u}K_f(r)$ for some (but not all) $r\in\tilde{\mathcal{Z}}_r$.
} that there exists a constant $c>0$, such that for every constraint $i=1,\dots,n_z$, we have either $\inf_{r\in\tilde{\mathcal{Z}}_r}L_{PK,i}(r)\geq c$ or $\sup_{r\in\tilde{\mathcal{Z}}_r}L_{PK,i}(r)=0$. 
For a $T$-periodic target signal $y^e$,  the resulting reference $r^*_{\cdot|t_i}$ converges to the optimal reachable trajectory $r^{T*}$ in finite time and the state $x_t$ converges exponentially fast to $x^{T*}_t$.  
\end{prop}
\begin{pf}
\textbf{Part I. } Recursive feasibility:
First, for $t=t_i+k$, $k=1,\dots,M-1$ the reference $r_{\cdot|t}$ satisfies the tightened constraints~\eqref{eq:limon_alpha_1_M} with $\alpha^{s*}_{t_i}$. 
Thus, feasibility of~\eqref{eq:MPC_M} at time $t_i$ implies recursive feasibility of~\eqref{eq:MPC_M} at $t$ with the updated terminal set size $\alpha^{tr}_t\leq \alpha^{s*}_{t_i}$ according to~\eqref{eq:alpha_s_update}, the standard MPC candidate solution from Theorem~\ref{thm:MPC} and the contractivity~\eqref{eq:contract}. 
Correspondingly, at time $t_{i+1}$, the candidate solution satisfies
\begin{align}
\label{eq:rho_M}
V_f(x_{N|t_{i+1}},r^*_{N+M|t_i})\leq \rho^{2M}\alpha^{tr}_{t_i}.
\end{align}
The constraint~\eqref{eq:alpha_s_M} ensures 
\begin{align*}
&\sqrt{V_f(x_{N|t_{i+1}},r^*_{N|t_{i+1}})}\\
\stackrel{\eqref{eq:term_cont}\eqref{eq:rho_M}}{\leq} &\rho^M\sqrt{\alpha^{tr}_{t_i}}(1+L_p\|r^*_{N+M|t_{i}}-r^*_{N|t_{i+1}}\|)\\
&+\sqrt{V_f(x^{r*}_{N+M|t_i},r^*_{N|t_{i+1}})}
\stackrel{\eqref{eq:alpha_s_M}}{\leq}  \alpha^{s*}_{t_{i+1}},
\end{align*}
which in combination with the update~\eqref{eq:alpha_s_update} and \eqref{eq:limon_alpha_bounds_M} implies $ \alpha^{tr}_{t_{i+1}}\leq \alpha^{s*}_{t_{i+1}}$. 
At time $t_{i+1}$ a feasible solution to~\eqref{eq:limon_alpha_M} is given by the  previous reference $r$ shifted by $M$ steps, i.e., $r_{j|t_{i+1}}=r^*_{\text{mod}(j+M,T)|t_i}$, $j=0,\dots,T-1$, with the candidate terminal set size 
\begin{align*}
\alpha^s_{t_{i+1}}=\max\{\rho^M\sqrt{\alpha^{tr}_{t_i}}+0.5(1-\rho^M)\sqrt{\alpha_ {\min}},\sqrt{\alpha_{\min}}\}
\end{align*}
and condition~\eqref{eq:alpha_s_M} strictly satisfied using the fact hat $V_f(x^{r*}_{N+M|t_{i}},r_{N|t_{i+1}})=0$  by definition. \\
\textbf{Part II. } Convergence: 
Consider the auxiliary candidate reference $\hat{r}$ based on $\hat{y}^r$ from \eqref{eq:y_r} with some $\beta_{t_i}\in[0,1]$. 
Suppose $\beta_{t_i}$ is chosen, such that $\|\hat{r}-r_{\cdot|t_{i+1}}\|\leq \epsilon$, with some constant $\epsilon$.
There exists a constant $\epsilon_1>0$, such that for $\epsilon\leq \epsilon_1$ this auxiliary reference satisfies the constraint~\eqref{eq:alpha_s_M}, with
\begin{align*}
&\rho^M\sqrt{\alpha^{tr}_{t_i}}(1+L_p\|r_{N|t_{i+1}}-\hat{r}_{N}\|)
+\sqrt{V_f(x^r_{N|t_{i+1}},\hat{r}_{N})}\\
\leq &\rho^M\sqrt{\alpha^{tr}_{t_i}}+(\sqrt{\alpha_1}\rho^ML_p+\sqrt{c_u})\epsilon_1\\
= &\rho^M\sqrt{\alpha^{tr}_{t_i}}+(1-\rho^M)0.5\sqrt{\alpha_{\min}}\leq \alpha^s_{t_{i+1}}.
\end{align*}
We show satisfaction of~\eqref{eq:limon_alpha_1_M} for the auxillary reference $\hat{r}$ with a case distinction. \\
\textbf{Case 1: } Suppose that $\alpha^s_{t_{i+1}}=\sqrt{\alpha_{\min}}$. 
In this case condition~\eqref{eq:limon_alpha_1_M} is equivalent to $\hat{r}\in\tilde{\mathcal{Z}}_r$, which is guaranteed by the convexity condition (Ass.~\ref{ass:unique_convex}) as in Prop.~\ref{prop:limon_alpha}.  \\
\textbf{Case 2: } $\alpha^s_{t_{i+1}}=\rho^M\sqrt{\alpha^{tr}_{t_i}}+0.5(1-\rho^M)\sqrt{\alpha_{\min}}$. 
Given $P_f$, $K_f$ and $P_f^{-1}$ continuous, $\mathcal{Z}$ compact and continuity of the quadratic norm, there exists a function $\delta\in\mathcal{K}_{\infty}$, such that for any $r,\tilde{r}\in\mathcal{Z}$:
\begin{align*}
&L_{PK,i}(r)-L_{PK,i}(\tilde{r})\leq \delta(\|r-\tilde{r}\|),~i=1,\dots,n_z. 
\end{align*}
For constraints $i$ with $L_{PK,i}(\hat{r}(j))=0$ feasibility of~\eqref{eq:limon_alpha_1_M} is independent of $\alpha^s$ and thus follows from convexity (Ass.~\ref{ass:unique_convex}). 
For the other constraints $i$ satisfaction of condition~\eqref{eq:limon_alpha_1_M} at $t_{i+1}$ follows from feasibility of~\eqref{eq:limon_alpha_1_M} at $t_i$ together with the definition of the candidate reference $r_{j|t_{i+1}}=r^*_{\text{mod}(j+M,T)|t_i}$, $\alpha^{tr}_{t_i}\leq \alpha^{s*}_{t_i}$ and $L_{PK,i}(\hat{r}_j)>0$:
\begin{align*}
&L_{PK,i}(\hat{r}_j)\alpha^s_{t_{i+1}}+L_{i}\hat{r}_j\\
\leq&L_{PK,i}(\hat{r}_j)
\left(\rho^M\sqrt{\alpha^{tr}_{t_{i}}}+0.5(1-\rho^M)\sqrt{\alpha_{\min}}\right)\\
&+L_{i}(\hat{r}_j-r_{j|t_{i+1}})+l_i
-L_{PK,i}({r}_{j|t_{i+1}})\sqrt{\alpha^{tr}_{t_i}}\\
\leq& l_i+\sqrt{\alpha_1}\delta(\|r_{\cdot|t_{i+1}}-\hat{r}\|)+\|L_i\|\|r_{\cdot|t_{i+1}}-\hat{r}\|\\
&-(1-\rho^M)\underbrace{L_{PK,i}(\hat{r}_j)}_{\geq c}\underbrace{(\sqrt{\alpha^{tr}_{t_i}}-0.5\sqrt{\alpha_{\min}})}_{\geq 0.5\sqrt{\alpha_{\min}}}
\leq l_i,
\end{align*}
where the last inequality holds for $\epsilon\leq \epsilon_2$ with some $\epsilon_2>0$. 
The reference satisfies
\begin{align*}
&\|r_{\cdot|t_{i+1}}-\hat{r}\|
\leq L_g\|\hat{y}^r-y^r_{\cdot|t_{i}}\|_S \\
\stackrel{\eqref{eq:1_beta}}{=}&(1-\beta_t)L_g\|y^{T*}_{\cdot|t_{i+1}}-y^r_{\cdot|t_{i+1}}\|_S,\nonumber
\end{align*}
with some Lipschitz constant $L_{g}$, compare Ass.~\ref{ass:unique_convex}. 
Thus, choosing $\beta_t=\max\{1-\epsilon/(L_g\|y^r_{\cdot|t_{i+1}}-y^{T*}_{\cdot|t_{i+1}}\|_S),0\}< 1$ with $\epsilon=\min\{\epsilon_1,\epsilon_2\}$, the candidate reference $\hat{r}$ is feasible. 
In case $\|y^r_{\cdot|t_{i+1}}-y^{T*}_{\cdot|t}\|_S\geq \epsilon/L_g$, this implies 
\begin{align*}
&J_T(\hat{r},t_{i+1})-J_T(r^*_{\cdot|t_i},t_i)\\
\stackrel{\eqref{eq:decrease_J_T} }{\leq}& -(1-\beta_t)(\epsilon/L_g)^2\leq -\epsilon^3/(L_g^3\sqrt{y_{\max}}),
\end{align*}
with $y_{\max}$ as defined on page~6.  
In case $\|y^r_{\cdot|t_{i+1}}-y^{T*}_{\cdot|t}\|_S\leq \epsilon/L_g$, the candidate reference converges to the optimal reference trajectory in one step, i.e., $\beta_t=0$, $J_T(\hat{r},t_{i+1})=V_T$. 
This ensures convergence of $J_T$ to $V_T$ and thus $r$ to $r^{T*}$ in at most $T_{\max}=M((\sqrt{y_{\max}}L_g/\epsilon)^3+1)$ time steps.
Exponential convergence of $x$ to $x^{T*}$ follows from exponential stability (Thm.~\ref{thm:MPC}) and finite time convergence of the reference $r$. 
$\hfill\square$
\end{pf}
The result in Proposition~\ref{prop:asynch} essential builds on three properties. 
First, the constraints on the reference planner~\eqref{eq:alpha_s_M} and the tracking MPC~\eqref{eq:MPC_M} with the contracting terminal set size~\eqref{eq:alpha_s_update} are such that the proposed algorithm is recursively feasible. 
Second, in each time step $t_{i}$, it is possible to incrementally reduce the tracking cost $J_T$, which in combination with convexity (Ass.~\ref{ass:unique_convex}) and compact constraints implies that the reference planner converges to the optimal reference $r^{T*}$ in finite time.  
Finally, once the reference planner converged in finite time, the tracking MPC~\eqref{eq:MPC_M} ensures exponential convergence to the reference $r=r^{T*}$ (c.f. Thm.~\ref{thm:MPC}) and thus exponential convergence of $x_t$ to $x^{T*}$. 
\begin{rem}
We point out that we only showed convergence for this partially coupled approach, as opposed to uniform stability in Thm.~\ref{thm:track}. 
It is possible to adjust Alg.~\ref{alg:asyn}, such that the reference planner does not require explicit information from the system, by replacing the update $\alpha^{tr}_t$ in~\eqref{eq:alpha_s_update} and just using the fact that the terminal set is $\rho$-contractive. 
Although this may simplify the computation, the closed-loop convergence of the tracking MPC is typically significantly faster, which is why the update~\eqref{eq:alpha_s_update} can speed up the convergence rate of the reference planner. 
Furthermore, it is possible to implement Alg.~\ref{alg:asyn} in an asynchronous fashion with $M$ changing online, if the constraint~\eqref{eq:alpha_s_M} is adjusted to hold for any $M\in[M_{\min},M_{\max}]\subset\mathbb{N}$. 
This way, the reference planner needs to solve~\eqref{eq:limon_alpha_M} until $t_i+M_{\max}$, but the reference can also be updated earlier starting at $t_i+M_{\min}$. 
\end{rem}
%
% 
%!TEX root = ./PeriodicTracking_Automatica.tex
%%%%%%%%%%%%%%%%%%%%%%%%%%%%%%%%%%%%%%%%%%%%%%%%%%%%%%%%%%%%%%%%%%%%%%%%%%%%%%%
\section{Numerical examples}
\label{sec:num}
The following examples show the general applicability of the proposed method and illustrate advantages compared to existing approaches. 
We first benchmark the performance of the proposed approach at the example of setpoint tracking of a continuous stirred tank reactor (CSTR).  
Then, we demonstrate the applicability of nonlinear periodic reference tracking at the example of a ball and plate system.  
The offline and online computation is done using SeDuMi-1.3~\cite{sturm1999using} and CasADi~\cite{andersson2019casadi}, respectively.
%!TEX root = ./PeriodicTracking_Automatica.tex
%%%%%%%%%%%%%%%%%%%%%%%%%%%%%%%%%%%%%%%%%%%%%%%%%%%%%%%%%%%%%%%%%%%%%%%%%%%%%%%
\subsection{Setpoint tracking - CSTR}
\label{sec:num_setpoint}
The following example demonstrates the performance benefits of the proposed method, especially the parameterized terminal ingredients (Lemma~\ref{lemma:LPV_term}) and the online optimized terminal set (Prop.~\ref{prop:limon_alpha}), at the example of set point tracking ($T=1$). \\
\textit{System model:} 
We consider a CSTR model\footnote{%
The parameters are $\theta_f=20,~k=300,~M=5,~x_f=0.3947,~x_c=0.3816,~\alpha_f=0.117$.
 } 
\begin{align*}
\dot{x}=
\begin{pmatrix}
\dot{x}_1\\\dot{x}_2
\end{pmatrix}
=
\begin{pmatrix}
\frac{1}{\theta_f}(1-x_1)-kx_1 e^{-\frac{M}{x_2}}\\
\frac{1}{\theta_f}(x_f-x_2)+kx_1 e^{-\frac{M}{x_2}}-\alpha_f u(x_2-x_c)
\end{pmatrix}
\end{align*}
with the concentration $x_1$, the temperature $x_2$ and the coolant flow rate $u$, taken from~\cite{mayne2011tube}. 
The discrete-time model is defined with an Euler discretization and the sampling time $h=0.1s$. 
We consider the setpoint tracking problem using the nonlinear MPC schemes~\eqref{eq:limon},~\eqref{eq:limon_alpha} in Section~\ref{sec:ext} with $T=1$ and the output $y=x_2$.  \\
\textit{Offline computations and terminal set:} 
The constraint set is given by $\mathcal{Z}=[0,1]^2\times[0,2]$, the stage cost is $Q=I_2$, $R=0.01$, the output weighting is $S=10^3$ and the feasible reference manifold is 
\begin{align*}
\mathcal{Z}_r=\{(x,u)|~f(x,u)=x,~x_2\in[0.43,0.86]\}.
\end{align*}
We compute the terminal ingredients (Lemma~\ref{lemma:LPV_term}) using \cite[Alg.~2]{JK_QINF} with a quasi-LPV parameterization $\theta\in\mathbb{R}^4$, gridding the steady-state manifold $\mathcal{Z}_r$ with $100$ points and $\epsilon=1$. 
The overall offline computations are accomplished in less than 10s. 

Regarding the terminal set size $\alpha$, we compare the fixed $\alpha$ based on $\mathcal{Z}_r$ with an online optimized, reference dependent, $\alpha(r)$ for the full constraint set $\mathcal{Z}$, compare~\eqref{eq:limon},~\eqref{eq:limon_alpha}.  
The resulting size of the terminal set for different setpoints $r$ can be seen in Fig.~\ref{fig:CSTR_setpoint_alpha}. 
We can see that the online optimization of $\alpha(r)$ (blue), allows us to consider states $x^r\notin\mathcal{Z}_r$ and thus yields a significantly larger operating area. 
For states $x^r\in\mathcal{Z}_r$, the constant terminal set size $\alpha$ (red) is considerably smaller and thus conservative. 
In~\cite{limon2018nonlinear}, it has been suggested to partition $\mathcal{Z}_r$ and compute different constants $\alpha_i$ for each partitioning. 
The terminal set size with $3$ equally spaced partitions of $\mathcal{Z}_r$ are also displayed in Fig.~\ref{fig:CSTR_setpoint_alpha}. 
The terminal sets based on partitioning is always an inner approximation to the continuously parameterized terminal sets with $\alpha(r)\geq \alpha_i$ for all $r\in\mathcal{Z}_r$. 
In addition, the continuously parameterized terminal ingredients are well suited for standard solvers with automatic differentiation, contrary to the piecewise constant definitions used in~\cite{limon2018nonlinear}. 

\begin{figure}[hbtp]
\begin{center}
\includegraphics[width=0.4\textwidth]{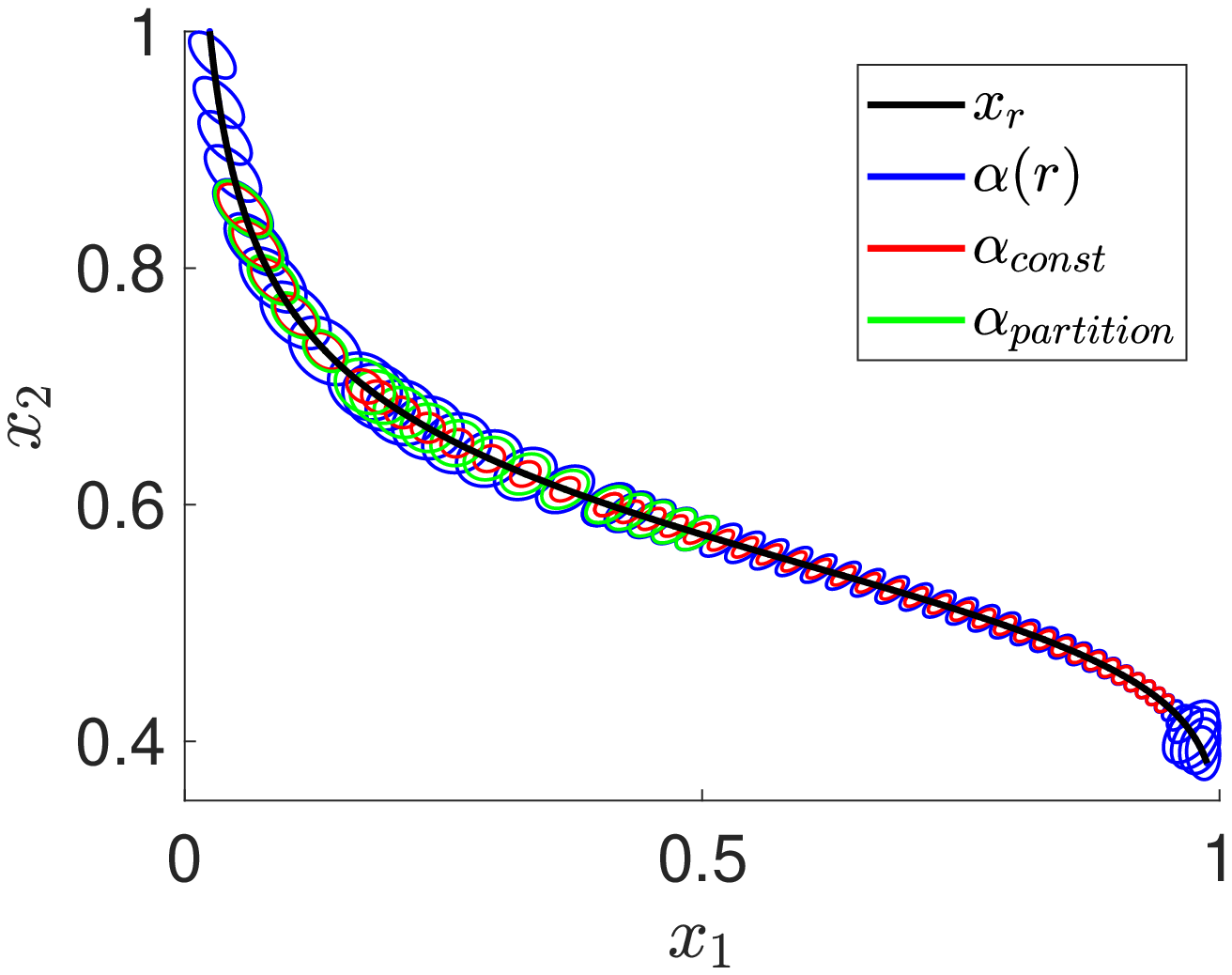}
\includegraphics[width=0.4\textwidth]{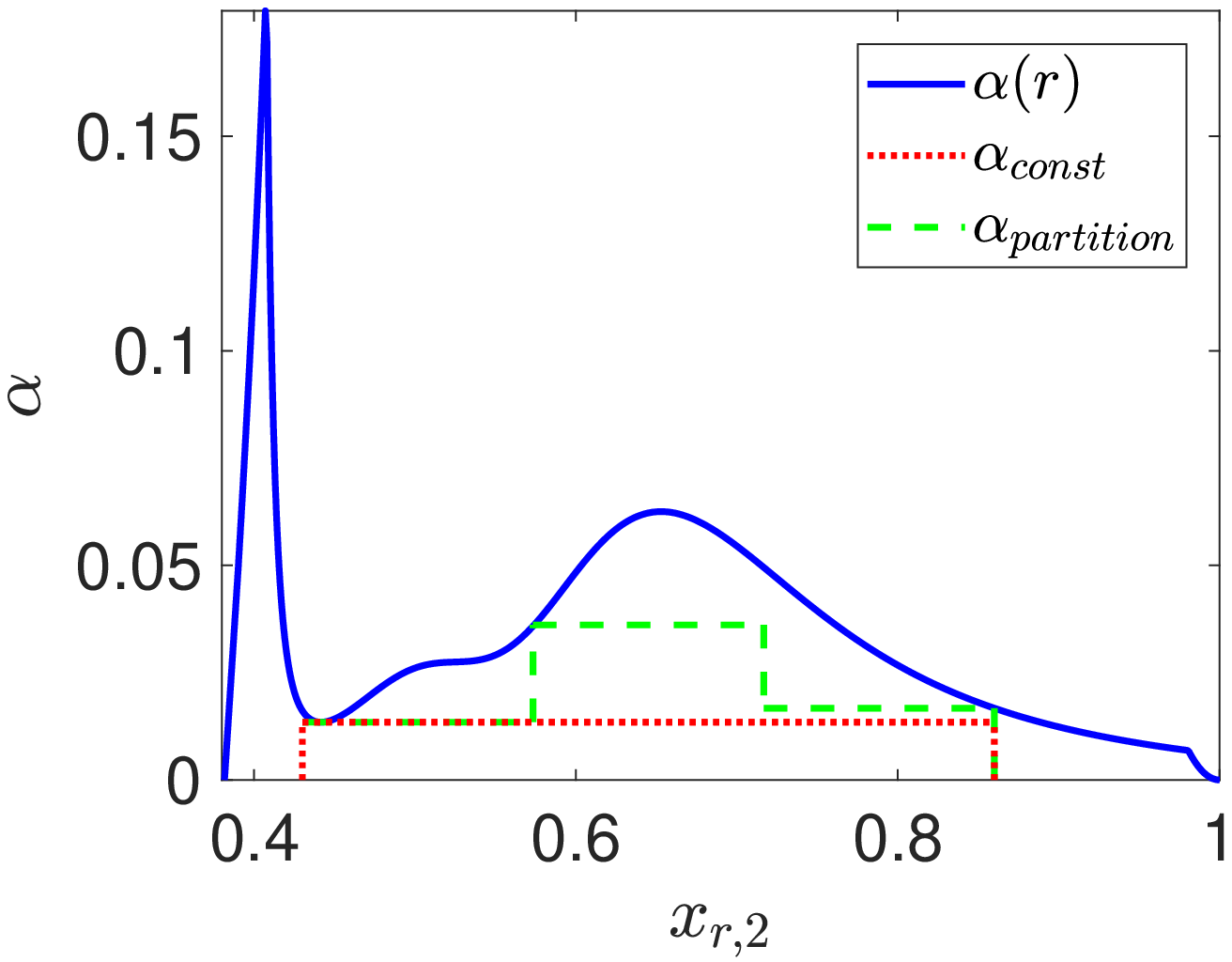}
\end{center}
\caption{Top: Temperature vs. concentration: setpoints $x^r$ (black) with corresponding size of terminal set $\mathcal{X}_f(r)$ for setpoint dependent $\alpha(r)$ (blue ellipses), constant $\alpha=0.013$ (red ellipses) and piece-wise constant $\alpha_{partition}$ (green ellipses). 
Bottom: Terminal set size $\alpha$ over setpoints $x_{r,2}$, for online optimized $\alpha(r)$ (solid blue), constant $\alpha$ (dotted red), piece-wise constant $\alpha_{partition}$ (dashed green).}
\label{fig:CSTR_setpoint_alpha}
\end{figure}
\textit{Setpoint tracking:} 
Starting at $x_0=[0.9492,0.43]$, the output target is $y^e=0.6519$, as in the numerical example in~\cite{mayne2011tube}. 
We implemented the proposed approach with terminal cost (Lemma~\ref{lemma:LPV_term}, QINF-$\alpha$) and with online optimized terminal set size (Prop.~\ref{prop:limon_alpha}, QINF-$\alpha(r)$). 
For comparison, we consider the terminal equality constraint tracking approach from~\cite{limon2018nonlinear} ($\mathcal{X}_f(r)=x^r$, Prop.~\ref{prop:TEC}, TEC)  as the state of the art solution. 
Furthermore, we implemented a modified version with the stage cost $\ell(x,u,r)=\|y-y^T\|^2$, similar to the \textit{economic} formulations in~\cite{fagiano2013generalized,muller2013economic} (TEC, \cite{fagiano2013generalized,muller2013economic}). 
In addition, we consider a reference governor\footnote{%
The local controller $k_f(x,r)$ (Ass.~\ref{ass:term_gen}) is applied.  
The reference $r$ is updated by increments of $x^{r+}_2=x^r_{},+0.003$, if $x\in\mathcal{X}_f(r^+)$ with the constant terminal set size $\alpha=0.013$. 
}, which corresponds to the candidate solution ($N=0$) in the stability proof (cf. Remark~\ref{rk:ref_govener}).

%\footnote{%
%The cumulative output tracking cost $\sum_{k=0}^{T}\|y_k-y^e_k\|^2$ is specified relative to the performance of the reference governor \JK{over a period of $T=5000$ steps}. The number of optimization variables are taken for the condensed formulation. 
%}
 Figure~\ref{fig:CSTR_setpoint_closedloop} shows a quantitative performance comparison and exemplary closed-loop trajectories for the different schemes. 
The different prediction horizons $N$ have been chosen, such that a similar performance is achieved, which again highlights the fact that the proposed approach can improve performance while simultaneously reducing the computational complexity, compared to state of the art formulations.
If we compare the performance of the proposed approach with suitably designed terminal ingredients (QINF) to simple terminal equality constraint (TEC) formulations, we can see a significant reduction of the tracking error even for $N=1$.  
Furthermore, we see that a suitably designed reference governor can compete with a badly designed (terminal equality constraint) tracking MPC (TEC, TEC~\cite{fagiano2013generalized,muller2013economic}), for short horizons $N$.
\begin{rem}
\label{rk:ballPlate_alternatives}
The closed-loop performance of the terminal equality constraint MPC (TEC) is very sensitive to the offset weighting, e.g. for $S=10^2$ the convergence rate decreases by one order of magnitude, while the proposed MPC with terminal cost is almost unaffected. 
Similarly, if we decrease the prediction horizon to $N=5$, the terminal equality constraint MPC has a significantly slower convergence by a factor of $25$. 
In~\cite[Sec. III.B]{limon2018nonlinear} it was suggested to drop the terminal set constraint and implicitly enforce the terminal constraint by scaling the terminal cost with some sufficiently large scaling factor $\gamma$.  
For the considered example this corresponds to $\gamma\approx 4\cdot 10^3$. 
With such a scaled terminal cost, the resulting closed-loop trajectory is virtually indistinguishable from the terminal equality constraint MPC and the ill-conditioning causes numerical difficulties. \\
An alternative approach to this problem would be to directly implement a stabilizing MPC without any terminal ingredients or artificial steady-state (`` unconstrained''), such as for example~\cite{boccia2014stability,kohlernonlinear19}. 
This scheme is quite sensitive to the stage cost $Q,~R$ and the considered prediction horizon $N$. 
For $N\leq 30$, the scheme simply gets stuck at a steady-state close to the initial state, while for $N=40$, the scheme shows fast convergence similar to the proposed scheme with terminal cost (QINF) and $N=15$ or terminal equality constraint (TEC) with $N=20$.  \\
If we use a stabilizing MPC with terminal ingredients (QINF) with a fixed artificial steady-state $r$, a prediction horizon of $N\geq 500$ is required to ensure initial feasibility, yielding a performance comparable to the proposed QINF-$\alpha(r)$ with $N=30$. 
Thus, using an artificial steady-state $r$ significantly reduces the computational demand and results in a smoother, potentially slower, closed-loop operation. \\
We point out that the alternative cost formulation from~\cite{fagiano2013generalized,muller2013economic} seems to yield better performance, given the same horizon and terminal ingredients (TEC vs. TEC~\cite{fagiano2013generalized,muller2013economic}).
Thus, combining the proposed approach with the \textit{economic} formulation in~\cite{fagiano2013generalized,muller2013economic} may yield even better performance, which is part of current research. 
\end{rem}
The proposed scheme uses both, the terminal ingredients computed offline (Lemma~\ref{lemma:LPV_term}) and online optimization. 
As a result, the scheme achieves superior performance with a small online computational demand.   
Furthermore, the benefits of optimizing the terminal set size $\alpha(r)$ online are clearly visible.   
The corresponding online optimization problem can be reduced to 2/3 scalar variables, thus ensuring real-time implementability. 
The considered example clearly demonstrates that a) the inclusion of suitable terminal ingredients is a major factor to ensure desired closed-loop performance, as articulated in~\cite{mayne2013apologia}, and b) the outlined procedure to compute terminal ingredients, including online optimization of the terminal set size (Sec.~\ref{sec:limon_alpha}), are well suited to improve the performance in nonlinear tracking MPC.
\begin{figure}[hbtp]
\begin{center}
\includegraphics[width=0.45\textwidth]{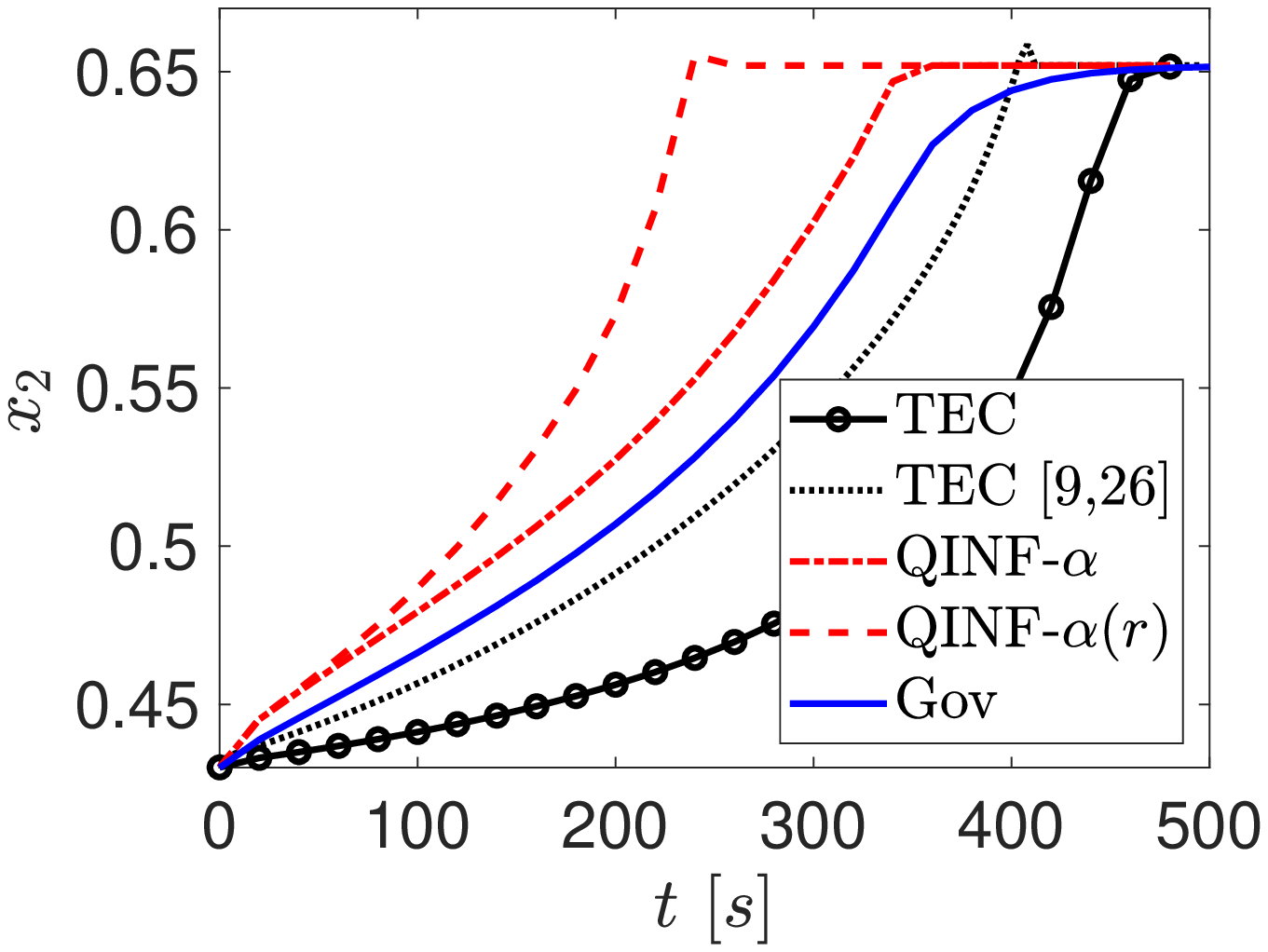}
\includegraphics[width=0.45\textwidth]{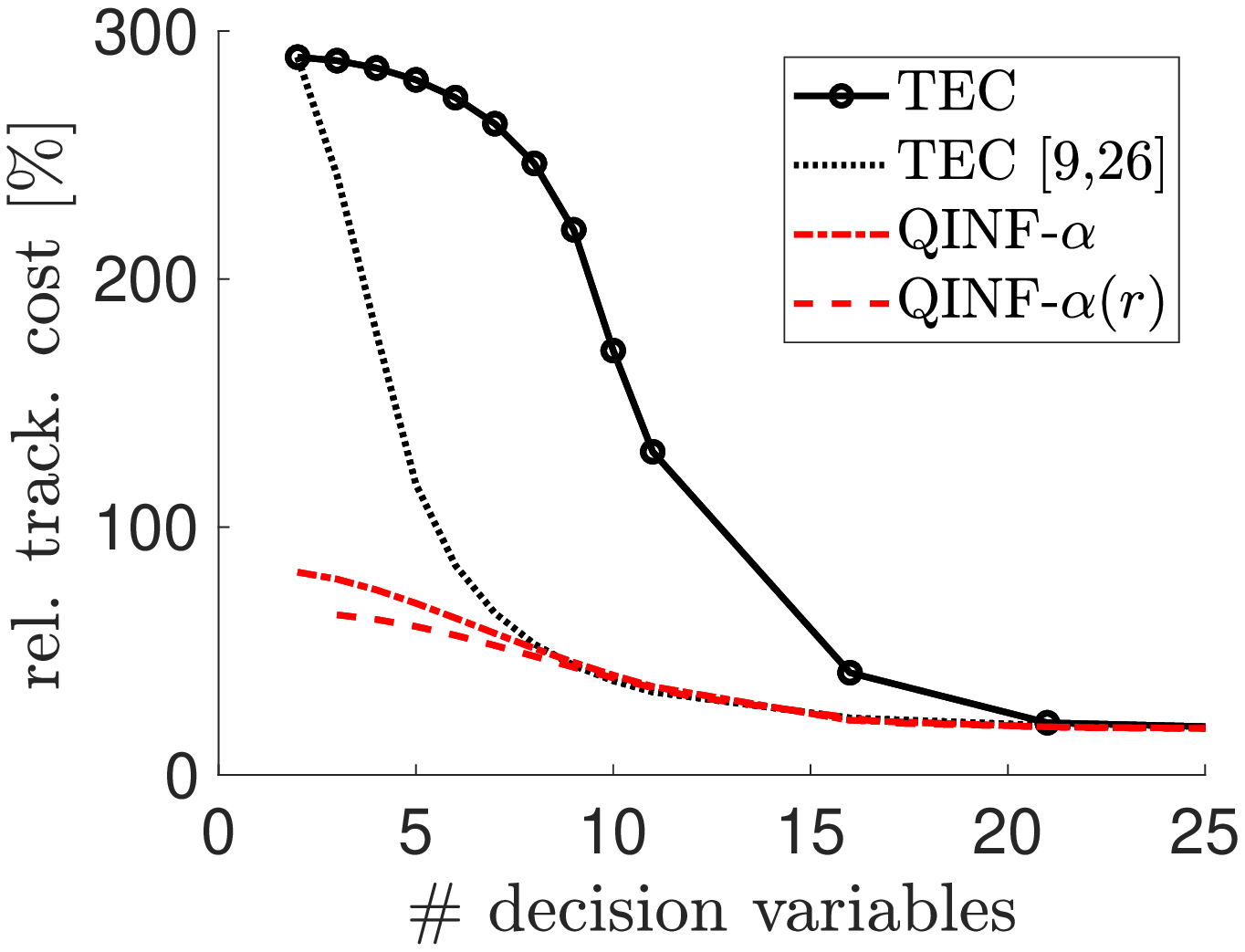}
\end{center}
\caption{Top: Exemplary closed loop (temperature vs. time): proposed scheme $N=1$ (red, QINF) with fixed $\alpha$ (dashed-dotted) and online optimized $\alpha(r)$ (dashed); reference governor (Gov, blue, solid); terminal equality constraint MPC $N=9$ (TEC, black, solid, circles), alternative cost $\ell$ with $N=4$ (TEC~\cite{fagiano2013generalized,muller2013economic}, black dotted). 
Bottom: Tracking cost $\sum_{t=0}^{5000}\|y_t-y^T_t\|^2$ relative to reference governor  v.s. number of decision variables (condensed formulation) in Problem~\eqref{eq:limon}.}
\label{fig:CSTR_setpoint_closedloop}
\end{figure}

%!TEX root = ./PeriodicTracking_Automatica.tex
%%%%%%%%%%%%%%%%%%%%%%%%%%%%%%%%%%%%%%%%%%%%%%%%%%%%%%%%%%%%%%%%%%%%%%%%%%%%%%%
\subsection{Periodic tracking - Ball and plate system}  
The following example shows the applicability of the proposed procedure to nonlinear periodic tracking. 
In addition, we demonstrate the practicality of the partially decoupled approach from Algorithm~\ref{alg:asyn}.  \\
\textit{System model:} 
We consider a nonlinear ball and plate system, taken from~\cite{pereira2017robust} with
\begin{align*}
\ddot{z}_1=&\frac{5}{7}(z_1\dot{\beta}_1^2+\dot{\beta}_1z_2\dot{\beta}_2+g\sin(\beta_1)),\\
\ddot{z}_2=&\frac{5}{7}(z_2\dot{\beta}_2^2+\dot{\beta}_2z_1\dot{\beta}_1+g\sin(\beta_2)),\\
x=&[z_1,z_2,\dot{z}_1,\dot{z}_2,\beta_1,\beta_2,\dot{\beta}_1,\dot{\beta}_2]^\top, \\
 u=&[\ddot{\beta}_1,\ddot{\beta}_2]^\top, \quad y=[z_1,z_2]^\top,
\end{align*}
with the position $z_i$ and the angle $\beta_i$.  
We use an Euler discretization of this model with step size $h=0.1s$ to get a nonlinear discrete-time system. \\
\textit{Offline computations:}
The constraint set is given by $\mathcal{Z}=[-0.06,0.06]^2\times[-0.2\times 0.2]^2\times[-\pi/3,\pi/3]^2\times[-1,1]^2\times [-2,2]^2$, the stage cost is $Q=I_8$, $R=0.1\cdot I_2$ and the output weighting is $S=I_2$. 
In contrast to the previous example, this system has a higher dimensionality $\mathcal{Z}\subset\mathbb{R}^{10}$ and a dynamic problem is considered. 
We compute constant\footnote{%
The Jacobian matrices $A(r)$, $B(r)$ are parameterized with $\theta\in\mathbb{R}^9$ (cf.~\cite{JK_QINF}), yielding $2^9=512$ vertices. 
If we wish to consider a larger constraint set, for example $|\dot{\beta}_i|\leq 2$, we cannot use constant matrices $P,K$, but reference dependent matrices $P(r),K(r)$ have to be computed. 
 In the numerical example in~\cite{kohler2018mpc} the matrices $P,~K$ were parameterized, such that $P$ only depends on $3$ parameters $\theta_i$. 
Thus, the offline computation had to consider $2^9\cdot 2^3=4096$ vertices, which increased the overall offline computation to approximately $1$~h.
} matrices $P,~K$ that satisfy the conditions in Lemma~\ref{lemma:LPV_term} with $\mathcal{Z}_r=\mathcal{Z}$ using~\cite[Alg.~2, Prop.~3]{JK_QINF} in $40~s$. \\
\textit{Periodic tracking:} In~\cite{limon2016mpc,pereira2017robust}, a linearized version of this model has been considered to study periodic reference tracking.  
Given the  reference generic terminal ingredients (Lemma~\ref{lemma:LPV_term}), we can extend these results to the nonlinear model with $T=16$. 
The initial condition and the target signal $y^e$ are chosen similar to~\cite{limon2016mpc,pereira2017robust}. 
In particular, the target signal $y^e$ is first an (unreachable) rectangular signal and suddenly changes to a circle with $T=16$. 
We implement the proposed approach with $N=1$ and online optimized terminal set size~\eqref{eq:limon_alpha}. 
In addition, we also implement the approach with the partially decoupled reference update (Alg.~\ref{alg:asyn}) with $N=1$, $M=2$. 
The resulting closed-loop trajectory can be seen in Figure~\ref{fig:Ball_closedloop}.
Initially, a large terminal set size $\alpha$ is optimal as it allows the controller to quickly move the reference but restricts the reference to have a large distance to the constraints. 
The reference then moves continuously to the optimal reachable trajectory $x^{T*}$ and the terminal set size $\alpha$ decreases to $\alpha_{\min}=10^{-8}$. 
As a result, the closed-loop trajectory shows initially fast convergence and then smoothly stabilizes the optimal trajectory $x^{T*}$.
The same effect can again be observed when the target signal suddenly changes at $t=5.5s$. 
The approach using partially decoupled reference updates has a slower convergence rate. 
On the other hand, this approach only uses $N\cdot m=2$ optimization variables to determine the control input $u$, while the reference update which requires $n+mT+1=41$ optimization variables, can be solved in intervals of $M\cdot h=200~ms$, thus greatly reducing the online computational demand. 
For comparison, the joint optimization~\eqref{eq:limon_alpha} requires $m(N+T)+n+1=43$ and needs to be solved every $h=100~ms$. 
The performance with different values of $M$ relative to the joint optimization~\eqref{eq:limon_alpha} is displayed in the following table.
\begin{tabular}{c|c|c|c|c|c}
$M$&$1$&$2$&$3$&$4$&$5$\\\hline
rel cost&$111\%$&$116\%$&$155\%$&$217\%$&$254\%$
\end{tabular}  
\begin{rem}
\label{rem:robust_TEC}
Since this numerical example has also been considered to study linear \textit{robust} tracking in~\cite{pereira2017robust}, we point out that the proposed scheme can be directly extended to ensure nonlinear robust tracking by including an appropriate constraint tightening, compare~\cite{Robust_TAC_19}, \cite[Thm.~2]{JK_QINF} and the discussion in Remark~\ref{rk:robust}. \\
Similar to the discussion in Remark~\ref{rk:ballPlate_alternatives}, one alternative to the proposed approach is to use an MPC scheme without any terminal ingredients or additional optimization variables and directly track the target signal $y^e$, compare~\cite{kohlernonlinear19}. 
Such an approach does not explicitly depend on the periodicity $T$ and requires no offline optimization, but typically needs a larger\footnote{%
In the considered example, such an approach yields reasonable results for $N\geq 13$, but  with a higher tracking error compared to the approach proposed in this paper.
} prediction horizon $N$.
\end{rem}
\begin{figure}[hbtp]
\begin{center}
\includegraphics[scale=0.5]{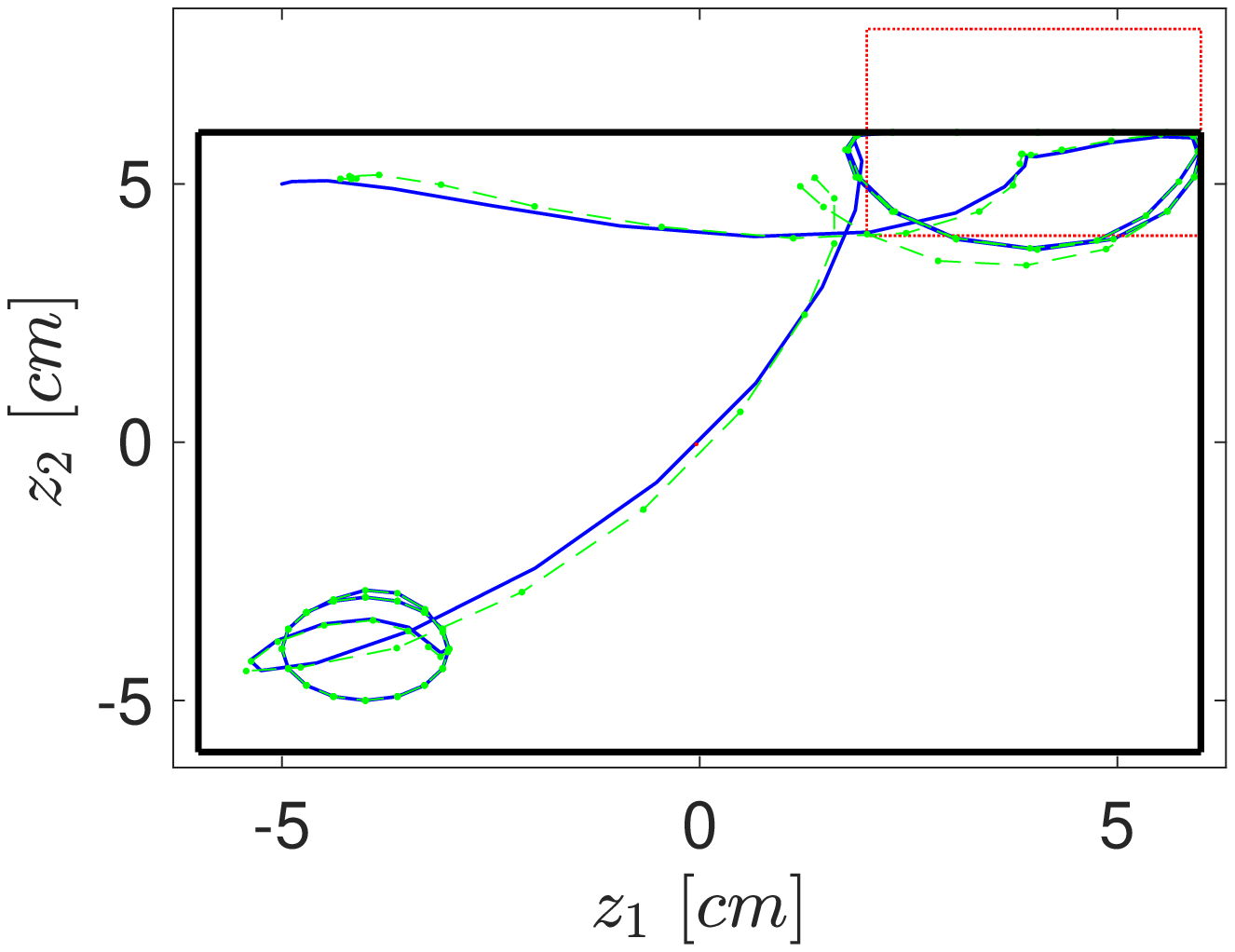}
\includegraphics[scale=0.5]{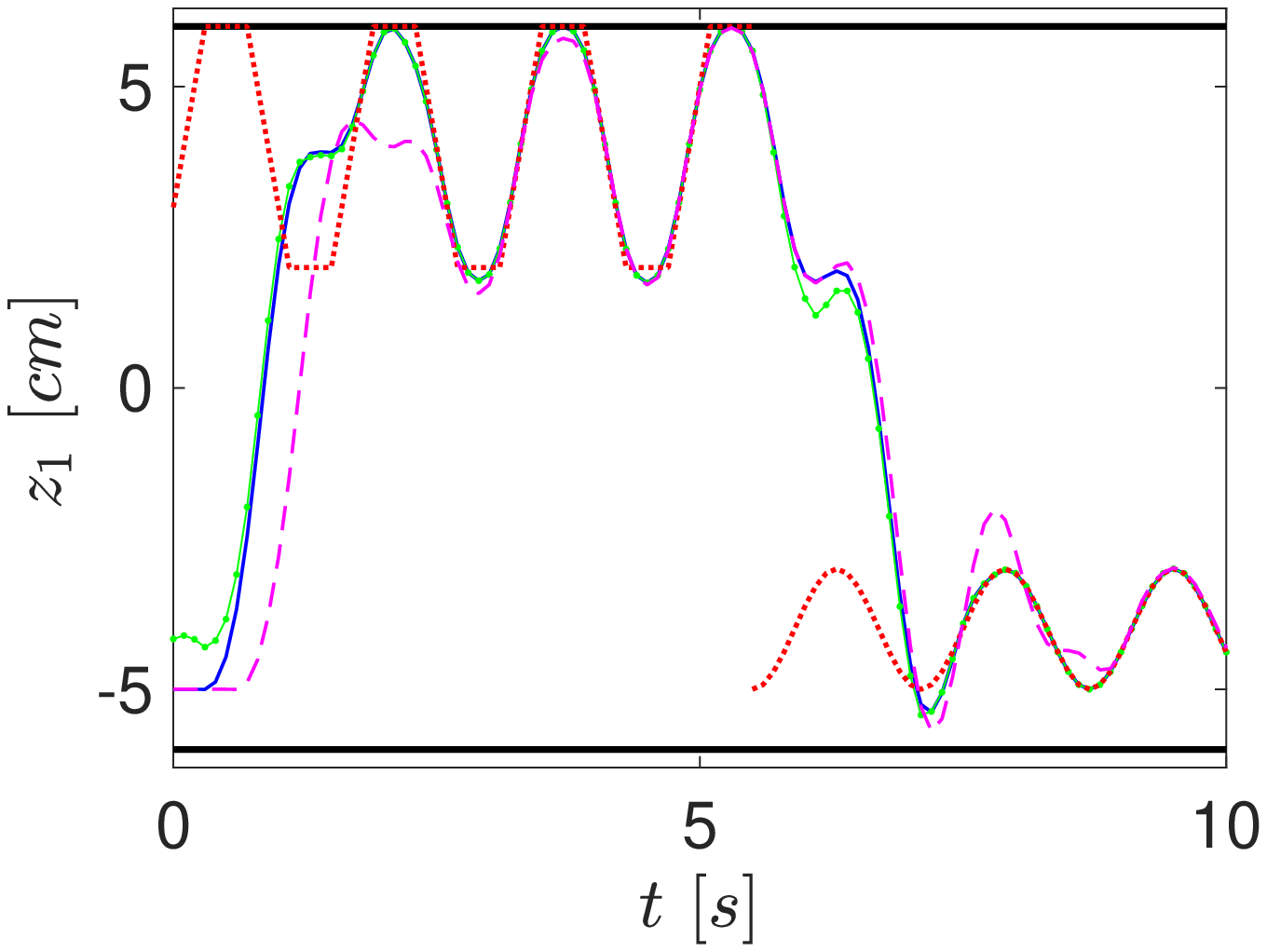}
\includegraphics[scale=0.5]{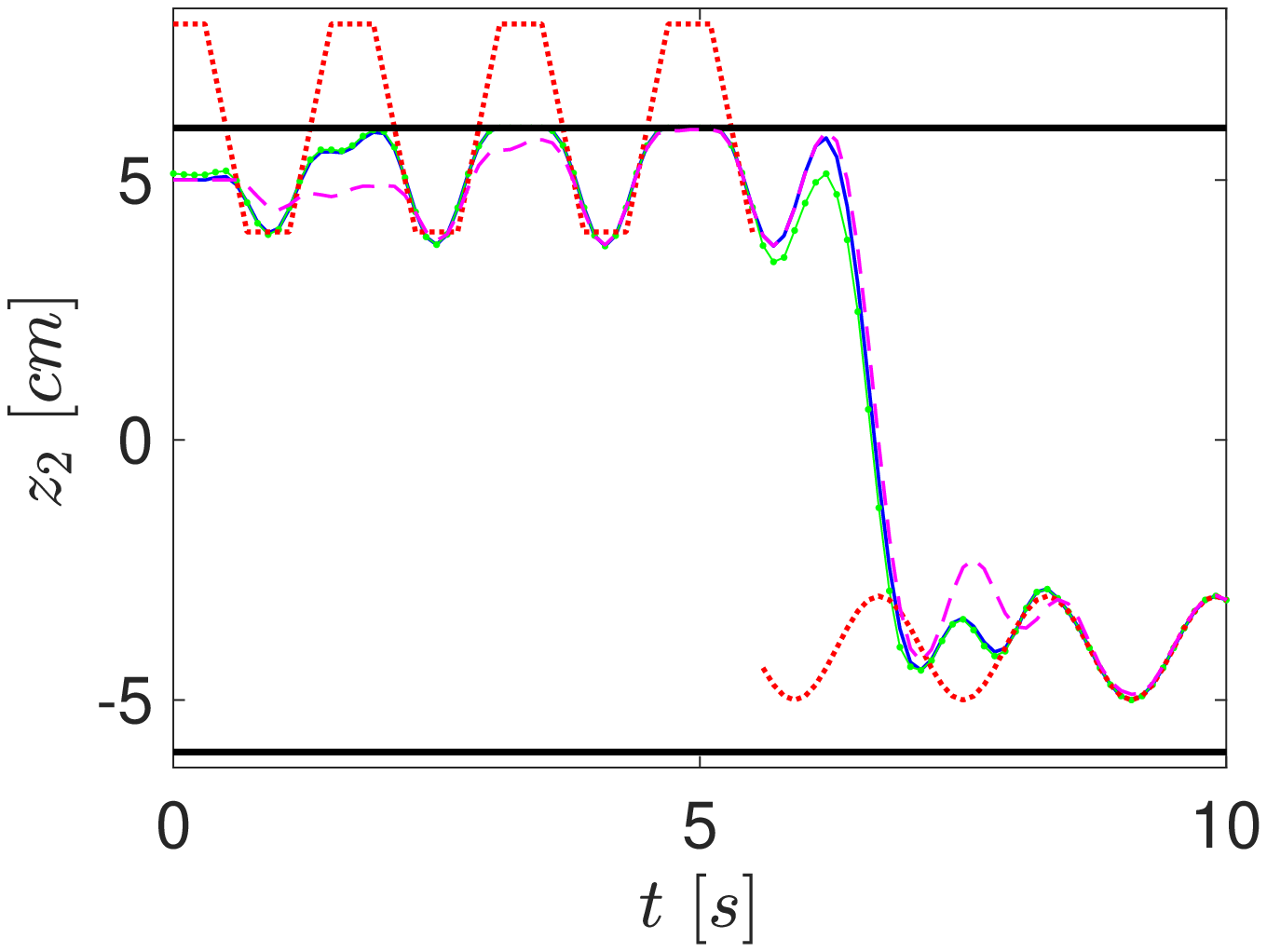}
\includegraphics[scale=0.45]{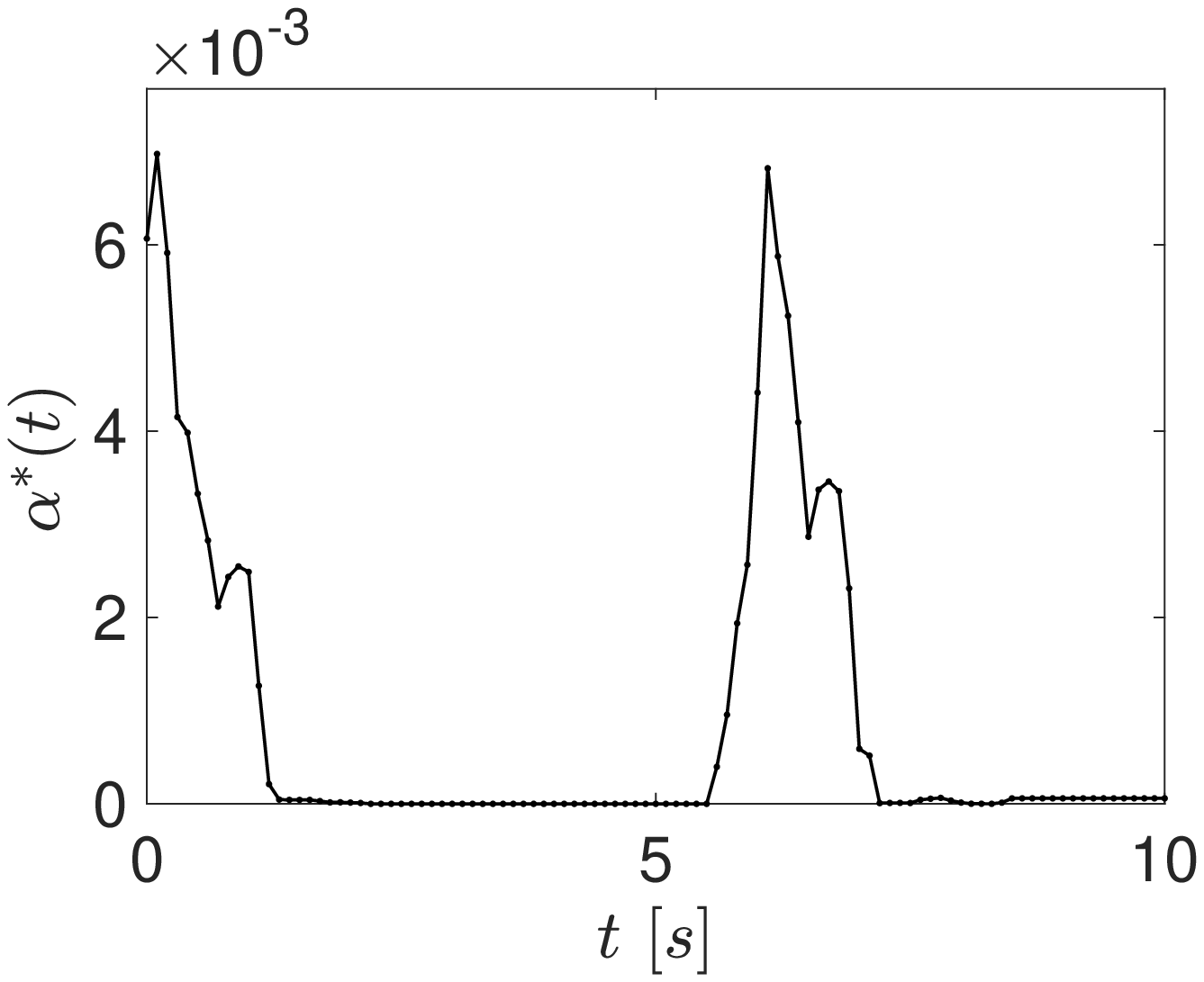}
\end{center}
\caption{Top: Trajectories of $z_1,~z_2$ for the closed-loop system $x$ (blue), the artificial reference $r$ (dash-dot green), the target signal $y^e$ (dotted red) and the state constraints (black). 
The closed-loop trajectories $z_1,~z_2$ for the closed-loop system based on partially decoupled updates (Alg.~\ref{alg:asyn}) (dashed magenta) is also displayed. 
Bottom: Closed-loop evolution of the online optimized terminal set size $\alpha$. }
 \label{fig:Ball_closedloop}
\end{figure}

%!TEX root = ./PeriodicTracking_Automatica.tex
%%%%%%%%%%%%%%%%%%%%%%%%%%%%%%%%%%%%%%%%%%%%%%%%%%%%%%%%%%%%%%%%%%%%%%%%%%%%%%%
\section{Conclusion}
\label{sec:sum}
We have proposed a nonlinear tracking MPC scheme for potentially unreachable periodic target signals using reference generic offline computations. 
If the set of periodic output trajectories is convex (Ass.~\ref{ass:unique_convex}), Theorem~\ref{thm:track} ensures exponential stability of the optimal (reachable) periodic trajectory. 
This result extends and unifies the results in~\cite{limon2018nonlinear,limon2016mpc} by considering \textit{nonlinear} systems, \textit{periodic} reference trajectories and considering different terminal ingredients (Ass.~\ref{ass:term_gen}).
We have extended this approach by introducing an online optimization of the terminal set size, which greatly improves the closed-loop performance in terms of region of operation and convergence speed. 
In addition, we have proposed a partially decoupled optimization problem that ensures recursive feasibility with a reduced computational demand. 
We have demonstrated the applicability of the proposed method using numerical examples. 
We have shown the benefits of the considered scheme including terminal cost, online optimized terminal set size and partially decoupled reference updates in a quantitative comparison to the state of the art approaces. 
%$
%
\begin{ack}                               
The authors thank the German Research Foundation (DFG) for support of this work within the Research Training Group Soft Tissue Robotics (GRK 2198/1 - 277536708).
\end{ack}

\bibliographystyle{plain}        
\bibliography{Literature}

\end{document}